\documentclass[aps,prd,twocolumn,tightenlines,preprintnumbers,showpacs,superscriptaddress,notitlepage,nofootinbib,floatfix,longbibliography]{revtex4-1}
\usepackage{graphicx}  
\usepackage{dcolumn}   
\usepackage{bm}        
\usepackage{amssymb}   
\usepackage{standalone}
\usepackage{enumitem}
\usepackage{xcolor}
\usepackage{slashed}
\usepackage{booktabs}
\usepackage{multirow}
\usepackage{amsmath}
\usepackage{stackrel}
\usepackage{rotating}
\usepackage{CJKutf8}
\usepackage{pifont}
\usepackage{mathtools}
\usepackage{hyperref}
\hypersetup{
    colorlinks=true,       
    linkcolor=blue,          
    citecolor=blue,        
    filecolor=blue,      
    urlcolor=blue           
}
\usepackage{simplewick}

\definecolor{kngrey}{HTML}{A6AAA9}
\definecolor{knred}{HTML}{EC5D57}
\definecolor{knorange}{HTML}{F39019}
\definecolor{knyellow}{HTML}{F5D328}
\definecolor{kngreen}{HTML}{70BF41}
\definecolor{knblue}{HTML}{51A7F9}
\definecolor{knpurple}{HTML}{B36AE2}

\def\mc#1{{\mathcal #1}}

\def\a{{\alpha}}
\def\b{{\beta}}

\def\D{{\Delta}}
\def\t{\tau}

\def\g{{\gamma}}
\def\G{{\Gamma}}

\def\l{{\lambda}}

\def\O{{\Omega}}

\def\mbfx{\mathbf{x}}

\def\eqref#1{{(\ref{#1})}}

\definecolor{gray}{rgb}{0.6,0.6,0.6}

\def\ip{{i^\prime}}
\def\jp{{j^\prime}}
\def\kp{{k^\prime}}
\def\ap{{\alpha^\prime}}
\def\bp{{\beta^\prime}}
\def\gp{{\gamma^\prime}}
\def\rp{{\rho^\prime}}

\begin{document}

\title{On the Feynman-Hellmann theorem in quantum field theory \\
and the calculation of matrix elements}

\author{Chris~Bouchard}
\email[]{chris.bouchard@glasgow.ac.uk}
\affiliation{School of Physics and Astronomy, University of Glasgow, Glasgow G12 8QQ, UK}
\affiliation{Department of Physics, The College of William \& Mary Williamsburg, VA 23187-8795, USA}

\author{Chia~Cheng~Chang \begin{CJK*}{UTF8}{bsmi}(張家丞)\end{CJK*}}
\email[]{chiachang@lbl.gov}
\affiliation{Nuclear Science Division, Lawrence Berkeley National Laboratory, Berkeley, CA 94720, USA}

\author{Thorsten~Kurth}
\email[]{tkurth@lbl.gov}
\affiliation{Nuclear Science Division, Lawrence Berkeley National Laboratory, Berkeley, CA 94720, USA}
\affiliation{NERSC, Lawrence Berkeley National Laboratory, Berkeley, CA 94720, USA}

\author{Kostas~Orginos}
\email[]{kostas@wm.edu}
\affiliation{Department of Physics, The College of William \& Mary Williamsburg, VA 23187-8795, USA}
\affiliation{Thomas Jefferson National Accelerator Facility Newport News, VA 23606, USA }                

\author{Andr\'{e}~Walker-Loud}
\email[]{awalker-loud@lbl.gov}
\affiliation{Nuclear Science Division, Lawrence Berkeley National Laboratory, Berkeley, CA 94720, USA}
\affiliation{Department of Physics,  The College of William \& Mary Williamsburg, VA 23187-8795, USA}
\affiliation{Thomas Jefferson National Accelerator Facility, Newport News, VA 23606, USA }                

\begin{abstract}
The Feynman-Hellmann theorem can be derived from the long Euclidean-time limit of correlation functions determined with functional derivatives of the partition function.
Using this insight, we fully develop an improved method for computing matrix elements of external currents utilizing only two-point correlation functions.
Our method applies to matrix elements of any external bilinear current, including nonzero momentum transfer, flavor-changing, and two or more current insertion matrix elements.
The ability to identify and control all the systematic uncertainties in the analysis of the correlation functions stems from the unique time dependence of the ground-state matrix elements and the fact that all excited states and contact terms are Euclidean-time dependent.
We demonstrate the utility of our method with a calculation of the nucleon axial charge using gradient-flowed domain-wall valence quarks on the $N_f=2+1+1$ MILC highly improved staggered quark ensemble with lattice spacing and pion mass of approximately 0.15~fm and 310~MeV respectively. We show full control over excited-state systematics with the new method and obtain a value of $g_A = 1.213(26)$ with a quark-mass-dependent renormalization coefficient.
\end{abstract}
\maketitle

\section{Introduction}
\label{Sec:1}
The Feynman-Hellmann theorem (FHT) in quantum mechanics relates matrix elements to variations in the spectrum~\cite{Guettinger1932, Pauli1933, Hellmann1933, PhysRev.56.340}:
\begin{equation}
\frac{\partial E_n}{\partial \l} = \langle n | H_\l | n \rangle\, ,
\end{equation}
where the Hamiltonian is given by $H = H_0 +\l H_\l$. This simple relation follows straightforwardly at first order in perturbation theory. The method is applicable beyond perturbation theory and is often used in lattice QCD (LQCD) calculations, for example, to compute the scalar quark matrix elements in the nucleon~\cite{Procura:2003ig,Procura:2006bj,Alexandrou:2008tn,WalkerLoud:2008bp,Ohki:2008ff,Young:2009zb,Durr:2011mp,Horsley:2011wr,Freeman:2012ry,Bali:2012qs,Oksuzian:2012rzb,Semke:2012gs,Shanahan:2012wh,Ren:2012aj,Junnarkar:2013ac,Durr:2015dna}
\begin{equation}\label{eq:qqbar_n}
	m_q \frac{\partial m_N}{\partial m_q} \bigg|_{m_q = m_q^\textrm{phy}}
= \langle \mathcal{N} | m_q \bar{q} q | \mathcal{N} \rangle\, ,
\end{equation}
for the light ($q=\{u,d\}$) and strange ($q=s$) quarks. Quantitative knowledge of these matrix elements is necessary for interpreting direct searches for dark matter which look for the elastic recoil of nuclei.  In the scenario that dark matter is heavy and couples through the electroweak sector, the uncertainty on the strange and charm nucleon matrix elements is one of the largest uncertainties in spin-independent constraints upon direct dark matter detection~\cite{Hill:2013hoa}. In particular, due to cancellations in the amplitude at the level of quarks and gluons, there is a particular sensitivity to the scalar charm quark matrix elements with current uncertainties allowing for several orders of magnitude variability in the cross section; see Fig.~3 of Ref.~\cite{Hill:2013hoa}. A significant reduction over the current uncertainty in these matrix elements would be a welcome advancement for the field.

Recently, the FHT has been used to compute other nucleon matrix elements, such as the spin content of the nucleon~\cite{Chambers:2014qaa,Chambers:2015bka}.
More recently, a hybrid method using ideas from background field methods~\cite{Fucito:1982ff,Martinelli:1982cb,Bernard:1982yu,Detmold:2006vu,Engelhardt:2007ub,Detmold:2009dx,Detmold:2010ts} and the FHT has been introduced to compute few-nucleon electroweak matrix elements~\cite{Savage:2016kon}.
An advantage of the FHT is that it relates a three-point correlation function to a change in a two-point correlation function induced by an external source.
Thus, one can take advantage of the simplified analyses of two-point functions.
Traditional lattice calculations of three-point functions, particularly those involving nucleons, face a number of challenging systematics beyond those present for two-point functions: the stochastic noise of three-point functions is more severe than the corresponding two-point functions and also three-point functions have systematic contamination from excited states which is constant in Euclidean time for fixed source-sink(insertion) separation with identical initial and final states at zero momentum transfer.  Controlling these systematics requires a significant increase in the numerical cost.

Previous implementations of the FHT and related methods~\cite{Chambers:2014qaa,Chambers:2015bka,Savage:2016kon} are also costly, as the calculation must be performed for several values of the external parameter, $\l$.
In the case of the scalar quark matrix elements, the QCD action contains the operators of interest, $\l = m_q$. The FHT is then simply used by varying the values of the quark masses and determining the resulting variation of the spectrum, a routine step in present LQCD calculations. In the case of the nucleon spin, the operator $\l\, \bar{q} \g_\mu \g_5 q$ is perturbatively added to the theory for varying values of $\l$ and the resulting spectrum is computed such that $\partial_\l E_n(\l)$ can be approximated via finite difference.

In this work, we develop an improved implementation of the FHT and explore its connection with the partition function of quantum field theory. This new method offers several advantages including: an improved implementation, improved stochastic sampling over computations of equal computing time, a complete discussion of all systematics, and demonstrably rigorous control over all systematics associated with analysis of correlation functions. To demonstrate these claims, we present the formulation of our method, and perform a sample calculation of the nucleon axial-vector charge.
We then discuss the generalizations and conclude.

\section{The Feynman-Hellmann Theorem and a new method \label{Sec:2}}

\subsection{The new method \label{sec:new_method}}

Consider a two-point correlation function computed in the presence of some external source
\begin{align}\label{eq:twopt_l}
C_\l(t) &= \langle \l | \mc{O}(t) \mc{O}^\dagger(0) | \l \rangle
\nonumber\\&
	= \frac{1}{\mc{Z}_\l} \int D \Phi e^{-S -S_\l}
		\mc{O}(t) \mc{O}^\dagger(0)
\end{align}
with the external source coupled through some bilinear current density $j(x)$
\begin{equation}\label{eq:action_l}
S_\l = \l \int d^4x j(x)\, ,
\end{equation}
and partition function in the presence of the source,
\begin{equation}\label{eq:Z_lam}
\mc{Z}_\l = \int D \Phi e^{-S -S_\l}.
\end{equation}
Here, $\Phi$ is a general field operator representing the various quantum fields of the theory. The state $|\l \rangle$ is the vacuum state in the presence of the external source. We denote the sourceless vacuum state, partition function, and two-point correlation function by 
\begin{align}
	|\Omega\rangle &= \lim_{\l\rightarrow0}|\l\rangle\, , \label{eq:vacuum_l} \\
	\mc{Z} &= \lim_{\l\rightarrow0} \mc{Z}_\l\, , \\
	C(t) &= \lim_{\l\rightarrow0} C_\l(t)\, ,
\end{align}
respectively. 
The operator $\mc{O}^\dagger(0)$ creates a tower of states with specified quantum numbers out of the vacuum at time $t=0$, which are later destroyed by a conjugate operator $\mc{O}(t)$ at time $t$.

We are interested in the partial derivative of this correlation function with respect to $\l$, at $\l = 0$. This partial derivative can be built from an integral of uniform functional derivatives over the space-time volume or, if we wish for more general matrix elements (such as those involving momentum transfer), an integral over nonuniform values of $\l(x)$.  For now, we will focus on the simplest case of a constant source, $\l(x) = \l$.

The partial derivative of interest is related to the matrix elements of the current $j(x)$
\begin{multline}\label{eq:dl_Z}
-\frac{\partial C_\l(t)}{\partial \l}\bigg|_{\l=0} =
	\frac{\partial \mc{Z}_\l}{\partial \l}\bigg|_{\l=0} \frac{ C(t) }{\mc{Z}}
\\	+ \frac{1}{\mc{Z}} \int D \Phi e^{-S} 
	\int d^4 x^\prime j(x^\prime)\ \mc{O}(t) \mc{O}^\dagger(0)\, .
\end{multline}
The first term is proportional to the vacuum matrix element of the current and vanishes unless the current has vacuum quantum numbers. The second term involves an integral over matrix elements involving the current and the creation/annihilation operators: 
\begin{align}\label{eq:dc_dl}
-\frac{\partial C_\l(t)}{\partial \l} \bigg|_{\l=0} =&\ 
	-C(t) \int dt^\prime \langle \O | \mathcal{J}(t^\prime)| \O \rangle 
\nonumber\\&
	+ \int dt^\prime \langle  \O | T\{ \mc{O}(t) \mathcal{J}(t^\prime) \mc{O}^\dagger(0) \} | \O \rangle
\end{align}
where we have defined $\mathcal{J}(t) = \int d^3 x j(t,\vec{x})$. The second term is related to the hadronic matrix of interest in the time region $0 < t^\prime < t$.  In the other time regions, $t^\prime < 0$ and $t^\prime > t$, the current $\mathcal{J}$ creates/destroys a tower of states that also couple to the states created by $\mc{O}$
(in the case of quark bilinear operators in QCD, these are just the mesons coupled to the $\bar{q}\,\G\,q$ currents):
\begin{align}\label{eq:dc_dl_noVac}
\int dt^\prime \langle  \O | &T\{ \mc{O}(t) \mathcal{J}(t^\prime) \mc{O}^\dagger(0) \} | \O \rangle
=
\nonumber\\&
	\phantom{+} \int_{-\infty}^0 dt^\prime \langle  \O | \mc{O}(t) \mc{O}^\dagger(0) \mathcal{J}(t^\prime) | \O \rangle
\nonumber\\&
	+\int_0^t dt^\prime \langle  \O | \mc{O}(t) \mathcal{J}(t^\prime) \mc{O}^\dagger(0) | \O \rangle
\nonumber\\&
	+ \int_t^\infty dt^\prime \langle  \O | \mathcal{J}(t^\prime) \mc{O}(t) \mc{O}^\dagger(0) | \O \rangle\, .
\end{align}

Recall that the FHT relates matrix elements to derivatives of the spectrum. In Euclidean calculations, the effective mass is a derived quantity which asymptotes to the ground-state energy in the long-time limit,
\begin{equation}
m^{\it eff}(t,\t) = \frac{1}{\t} \ln \left( \frac{C(t)}{C(t+\tau)} \right) \underset{t\rightarrow\infty}{\longrightarrow} \frac{1}{\tau} \ln (e^{E_0 \t}).
\label{eq:meff}
\end{equation}
In analogy with the FHT, consider the linear response of the effective mass to the external current
\begin{align}\label{eq:dm_dl}
	\frac{\partial m^{\it eff}_\l(t,\t)}{\partial \l} \bigg|_{\l=0}
	\kern -0.15cm = 
	\frac{1}{\t} \left[ 
		\frac{\partial_\l C_\l(t)}{C(t)} - \frac{\partial_\l C_\l(t+\t)}{C(t+\t)}  
	\right]_{\l=0} \kern -0.1cm .
\end{align}

A first observation to make is that the term proportional to the vacuum matrix element in Eq.~\eqref{eq:dc_dl} exactly cancels in the difference in Eq.~\eqref{eq:dm_dl}. The linear response of the effective mass is therefore given by
\begin{equation}\label{eq:dmeff}
\frac{\partial m^{\it eff}_\l(t,\t)}{\partial \l} \bigg|_{\l=0}
	= \frac{R(t+\t) - R(t)}{\t}
\end{equation}
where
\begin{equation}\label{eq:R(t)}
R(t) \equiv \frac{\int dt^\prime \langle \O | T\{ \mc{O}(t) \mathcal{J}(t^\prime) \mc{O}^\dagger(0) \} | \O \rangle}{C(t)}\, .
\end{equation} 

This expression can be analyzed with the usual spectral decomposition.
The two-point correlation function in time-momentum space, with $\mathbf{p}=\mathbf{0}$, is given by
\begin{align}\label{eq:2ptansatz}
C(t,\mathbf{0}) &= \sum_\mbfx 
	\langle \O | \mc{O}(t,\mbfx) \mc{O}^\dagger(0,\mathbf{0}) | \O \rangle
\nonumber\\&
	= \sum_n \frac{Z_{n}^{\mathbf{0}} Z^\dagger_n}{2E_n} e^{-E_n t}
\end{align}
which can be obtained by inserting the identity operator
\begin{equation}
1 = |\O\rangle \langle \O| + \sum_n \int \frac{d^3p}{2E_{n}(p)} 
	|n,\mathbf{p}\rangle\langle n,\mathbf{p}|\, .
\end{equation}
The overlap factors are defined as
\begin{align}
Z^\dagger_n &= \langle n| \mc{O}^\dagger(0,\mathbf{0}) | \O \rangle\, ,
\nonumber\\
Z_{n}^{\mathbf{p}} &= \sum_\mbfx e^{i\mathbf{p}\cdot\mbfx} 
	\langle \O | \mc{O}(0,\mbfx) | n \rangle\, .
\end{align}

\begin{figure*}
\begin{tabular}{cccc}
\includegraphics[width=0.22\textwidth]{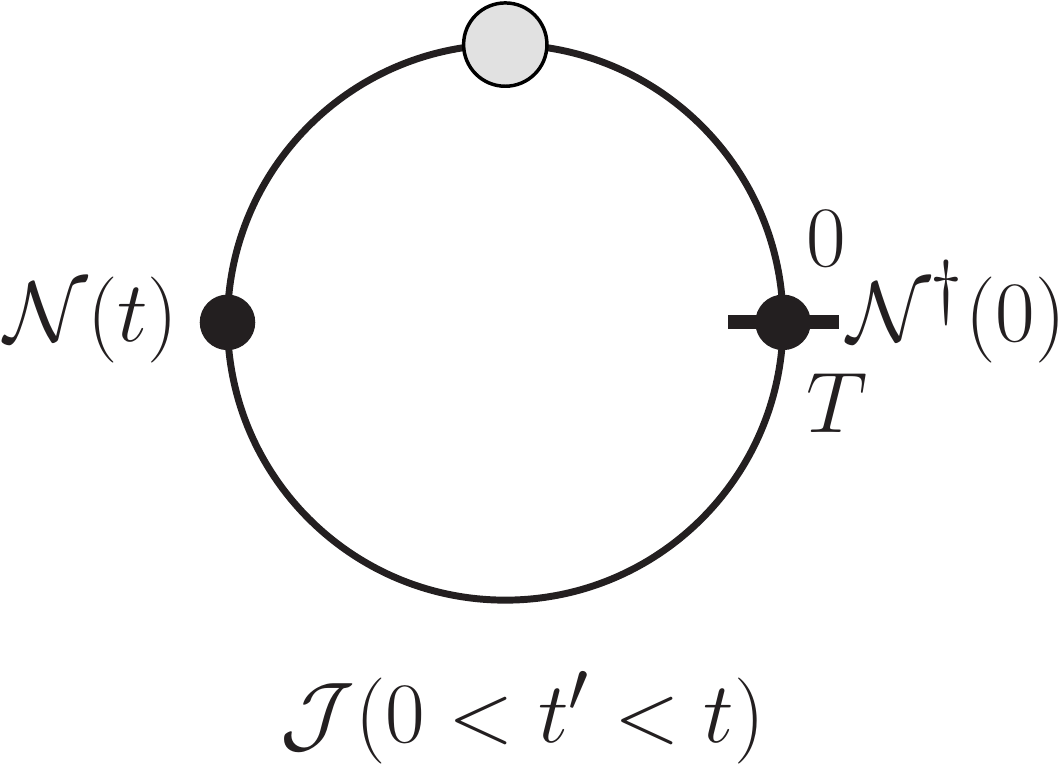}
&\includegraphics[width=0.22\textwidth]{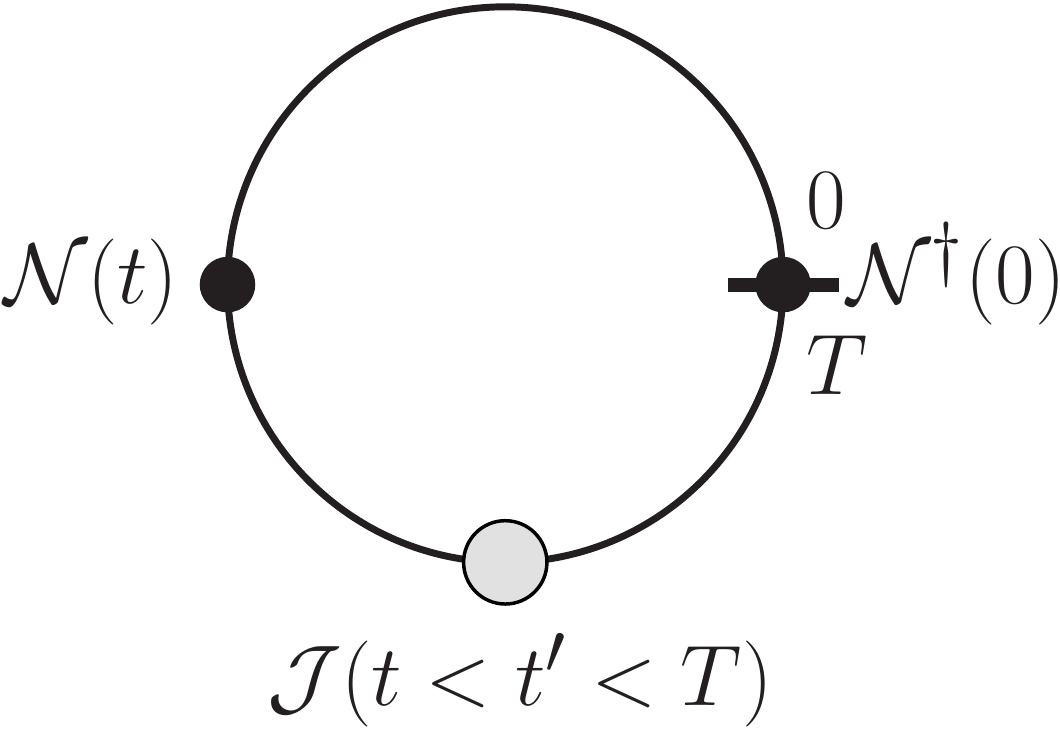}
&\includegraphics[width=0.22\textwidth]{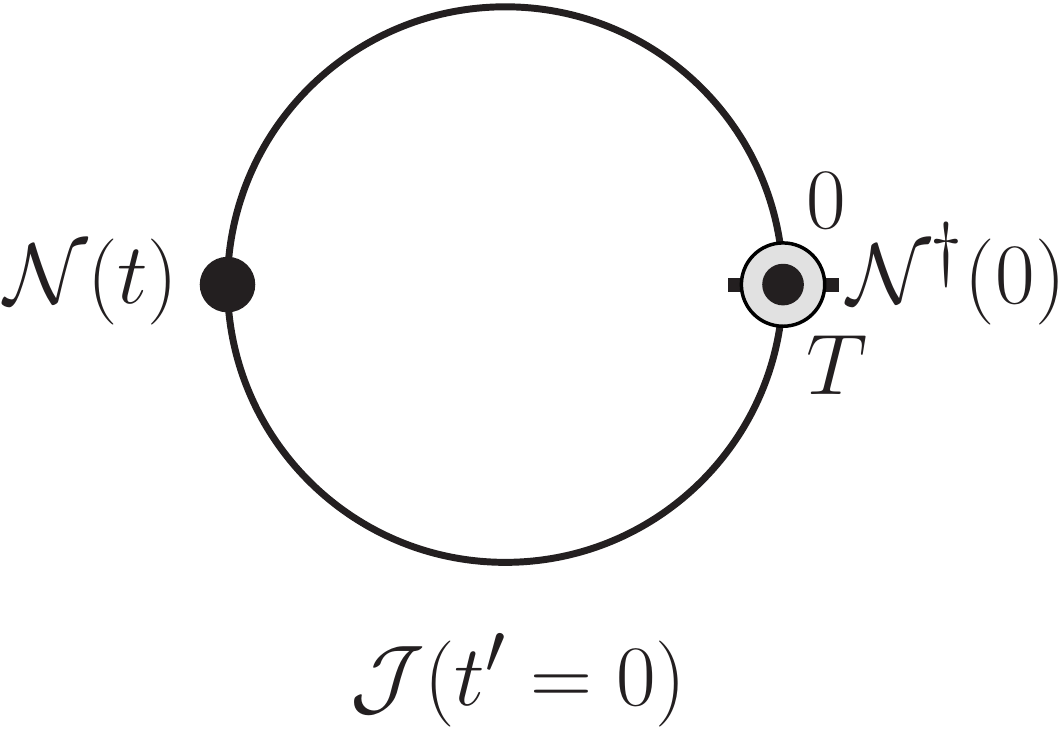}
&\includegraphics[width=0.22\textwidth]{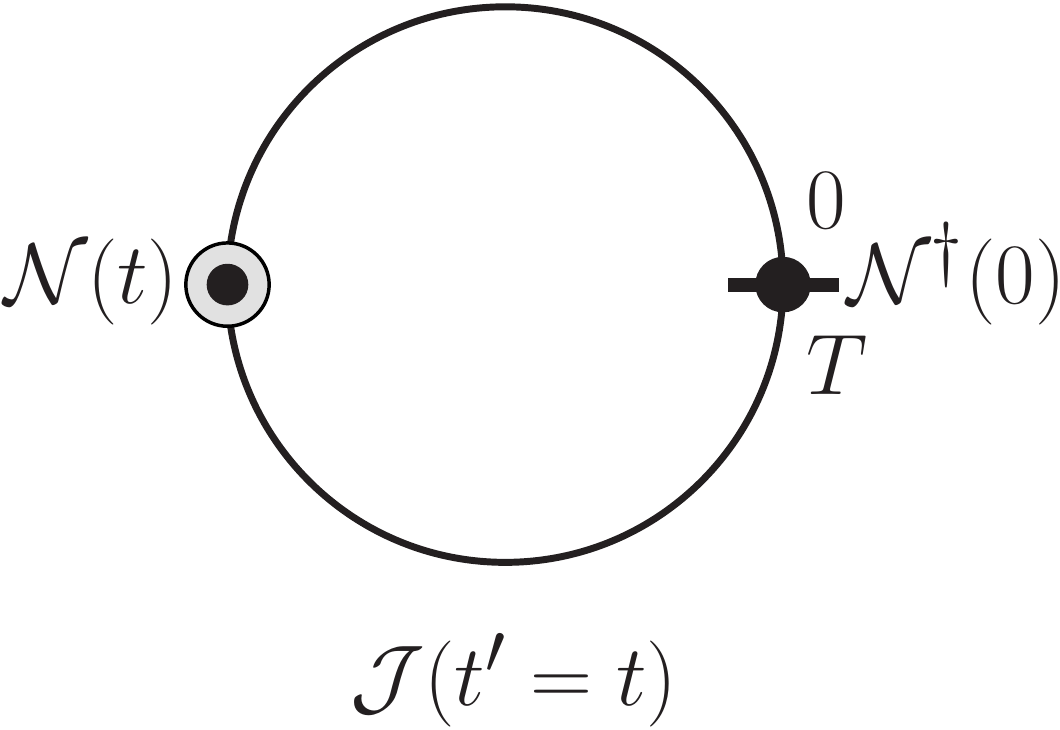}
\\
(I)& (II) &(III)& (IV)\\
\end{tabular}
\caption{\label{fig:fh_times} The four different current insertion time regions.
The matrix element of interest occurs in region I: $0<t^\prime<t$.  Regions II, III and IV are systematic corrections which must be accounted for where III and IV arise from contact operators.  The horizontal black line represents the temporal boundary with (anti)periodic boundary conditions. The solid black circles represent the hadron creation/annihilation operators and the light grey circle is the current insertion.}
\end{figure*}

The numerator in Eq.~\eqref{eq:R(t)} can be similarly decomposed.
It is useful to work in discrete Euclidean time relevant to LQCD calculations in which the numerator is
\begin{equation}
N(t) = \sum_{t^\prime=0}^{T-1} \langle 0 | T\{ \mc{O}(t) \mathcal{J}(t^\prime) \mc{O}^\dagger(0) \} | 0 \rangle\, .
\end{equation}
There are four time regions to consider, depicted in Fig.~\ref{fig:fh_times}:
\begin{align}
&\textrm{I: } 0 < t^\prime < t\, ,&
&\textrm{II: } t < t^\prime < T\, ,&
\nonumber\\
&\textrm{III: } t^\prime = 0\, ,&
&\textrm{IV: } t^\prime = t\, .&
\end{align}
The matrix elements of interest occur in time region I, when the current is inserted between the source/sink creation/annihilation interpolating fields.
The contributions from regions II, III and IV are systematic corrections which must be accounted for and controlled.
In region II, the current creates/destroys a meson with the quantum numbers of the current outside the region of interest.
Regions III and IV give rise to \textit{contact operators} when the current insertion time occurs precisely at the creation or annihilation time of the hadron.

In the first region, $0 < t^\prime < t$, the numerator is
\begin{equation}
N_\textrm{I}(t) = \sum_{t^\prime=1}^{t-1} \sum_{\b} \frac{e^{-E_\b T}}{\mc{A}_\b}
	\langle \b| \mc{O}(t) \mathcal{J}(t^\prime) \mc{O}^\dagger(0) | \b \rangle\, ,
\end{equation}
where $\mc{A}_\b=1$ for the vacuum and $\mc{A}_\b= 2E_\b$ otherwise.
All thermal states will be exponentially suppressed and we can consider the contribution from just the vacuum state $|\b\rangle = |\O\rangle$.  In this case, we have
\begin{align}\label{eq:N1_intermediate}
N_\textrm{I}(t) &= \sum_{t^\prime=1}^{t-1} \sum_{n,m}
	\langle \O | \mc{O}(t) \frac{|n\rangle \langle n|}{2E_n}
	\mathcal{J}(t^\prime) \frac{|m\rangle \langle m|}{2E_m}
	\mc{O}^\dagger(0) | \O \rangle
\nonumber\\&
	= \sum_{t^\prime=1}^{t-1} \sum_{n,m} 
		\frac{Z_{n}^{\mathbf{0}} Z^\dagger_m}{4E_n E_m}
		\langle n| \mathcal{J} | m \rangle e^{-E_n t} e^{-(E_m -E_n)t^\prime}
\nonumber\\&
	= \sum_{t^\prime=1}^{t-1}\bigg[ 
		\sum_n \frac{Z_{n}^{\mathbf{0}} Z^\dagger_n}{4E_n^2}
		\langle n| \mathcal{J} | n \rangle e^{-E_n t}
\nonumber\\&\
	+\sum_{n\neq m}
		\frac{Z_{n}^{\mathbf{0}} Z^\dagger_m}{4E_n E_m}
		\langle n | \mathcal{J} | m \rangle e^{-E_n t} e^{-(E_m - E_n) t^\prime}
	\bigg]\, .
\end{align}
The first term is independent of $t^\prime$ and is thus enhanced by an overall factor of $t-1$.
The $t^\prime$ dependence of the second term is contained entirely in the exponential factor
\begin{align}
\sum_{t^\prime=1}^{t-1} e^{-(E_m - E_n)t^\prime} = 
	\frac{1 - e^{-\D_{mn}(t-1)}}{e^{\D_{mn}} -1}\, ,
\end{align}
where we have defined
\begin{equation}
\D_{mn} \equiv E_m - E_n\, .
\end{equation}
The contributions from region I are then
\begin{multline}\label{eq:NI}
N_\textrm{I}(t) = (t-1) \sum_n \frac{Z_{n}^{\mathbf{0}} Z^\dagger_n}{4E_n^2}
	e^{-E_n t} J_{nn} 
\\
	+ \sum_{n,m\neq n} 
	\frac{Z_{n}^{\mathbf{0}} Z^\dagger_m}{4E_n E_m}
	\frac{e^{-E_n t -\frac{\D_{mn}}{2}} - e^{-E_m t -\frac{\D_{nm}}{2}}}
		{e^{\frac{\D_{mn}}{2}} -e^{\frac{\D_{nm}}{2}}}
	J_{nm} \, ,
\end{multline}
where we have defined $J_{nm} \equiv \langle n | \mathcal{J} | m \rangle$ and written the expression to expose the $n\leftrightarrow m$ symmetry.
The only terms which do not also appear in the two-point correlation functions are the matrix elements of interest, $J_{nm}$.
It is worth noting that the entire contribution from region I vanishes at $t=1$.

Let us now consider the contribution from region II,
\begin{equation}
N_\textrm{II}(t) = \sum_{t^\prime=t+1}^{T-1} \sum_{\b} \frac{e^{-E_\b T}}{\mc{A}_\b}
	\langle \b| \mathcal{J}(t^\prime) \mc{O}(t) \mc{O}^\dagger(0) | \b \rangle\, .
\end{equation}
To understand the systematics from this region, we cannot immediately drop the nonvacuum contributions from the thermal sum. Inserting a complete set of states, one finds
\begin{align}
N_\textrm{II}(t) &= \sum_{\b,\a,n} J_{\b\a} 
	\langle \a| \mc{O} |n\rangle \langle n| \mc{O}^\dagger |\b\rangle
	\frac{e^{-E_\b T} e^{E_\a t} e^{-E_n t}}{\mc{A}_\a \mc{A}_\b \mc{A}_n}
\nonumber\\& \quad
	\times \sum_{t^\prime=t+1}^{T-1} e^{t^\prime \D_{\b\a}}
\nonumber\\&=
	\sum_{\b,\a,n} 
	\frac{J_{\b\a} \langle \a| \mc{O} |n\rangle \langle n| \mc{O}^\dagger |\b\rangle}
		{\mc{A}_\a \mc{A}_\b \mc{A}_n} 
	e^{-E_n t}
\nonumber\\& \quad
	\times \frac{e^{-E_\a (T-t)}e^{\D_{\a\b}/2} - e^{-E_\b (T-t)} e^{\D_{\b\a}/2}}
		{e^{\D_{\b\a}/2} - e^{\D_{\a\b} / 2}}\, .
\end{align}
For small $t$, we can neglect all but the vacuum contribution in the sum over $\a$ and $\b$ for the first and second terms respectively. The region II contributions are then
\begin{multline}\label{eq:NII}
N_\textrm{II}(t) = \sum_{n} \frac{e^{-E_n t}}{2E_n} \bigg\{
	 Z_{n}^{\mathbf{0}} Z^\dagger_n \langle \O| \mathcal{J} |\O\rangle 	
\\
	 +\sum_{j} \frac{Z^{\mathbf{0}}_n Z^\dagger_{n j} J^\dagger_{j} 
		+ J_{j} Z_{j n}^{\mathbf{0}} Z^\dagger_n}{2E_j(e^{E_j} - 1)}
	\bigg\}
\end{multline}
where we have defined 
\begin{align}
J^\dagger_j \equiv \langle j| \mathcal{J} |\O\rangle\, ,&&
Z_{nj}^\dagger \equiv \langle n | \mc{O}^\dagger | j \rangle \, .
\end{align}
Note the first term in Eq.~\eqref{eq:NII} only contributes for scalar currents.

Finally, there is the contribution from the contact terms, regions III and IV, when $t^\prime=0$ or $t^\prime=t$. These contact contributions are standard two-point functions with different interpolating operators
\begin{align}
N_\textrm{III}(t) &= \sum_{\b,n} \frac{e^{-E_\b T}}{\mc{A}_\b}
	\langle \b| \mc{O}(t) \frac{|n\rangle \langle n|}{2E_n} \mathcal{J}(0) \mc{O}^\dagger(0) | \b \rangle\, ,
\\
N_\textrm{IV}(t) &= \sum_{\b,n} \frac{e^{-E_\b T}}{\mc{A}_\b}
	\langle \b| \mc{O}(t) \mathcal{J}(t) \frac{|n\rangle \langle n|}{2E_n} \mc{O}^\dagger(0) | \b \rangle\, .
\end{align}
In both terms, the thermal contributions are suppressed for all but the vacuum contribution. These two terms then contribute 
\begin{align}\label{eq:NIII_IV}
N_\textrm{III+IV}(t) &= \sum_{n} \frac{e^{-E_n t}}{2E_n}
	\left[ Z_n^\mathbf{0} Z^\dagger_{J:n}
	+ Z_{J:n}^\mathbf{0} Z^\dagger_n \right]\, ,
\end{align}
where we have defined
\begin{equation}
Z_{J:n}^\dagger \equiv \langle n | \mathcal{J} \mc{O}^\dagger | \O \rangle \, .
\end{equation}
Because the states $|n\rangle$ are annihilated by the same operator as in the two-point function, the sum over states in Eq.~\eqref{eq:NIII_IV} is over the same set of states as the two-point function but with modified overlap factors.

Putting all the regions together, the numerator is
\begin{multline}
N(t) = \sum_n \bigg\{ \frac{e^{-E_n t}}{2E_n} \bigg[
	(t-1)J_{nn} \frac{Z_{n}^{\mathbf{0}} Z^\dagger_n}{2E_n}
\\
	+Z_n^\mathbf{0} Z^\dagger_{J:n}
	+ Z_{J:n}^\mathbf{0} Z^\dagger_n 
	+Z_{n}^{\mathbf{0}} Z^\dagger_n \langle \O| \mathcal{J} | \O\rangle
\\
	+\sum_{j} \frac{Z^{\mathbf{0}}_n Z^\dagger_{n j} J^\dagger_{j} 
		+ J_{j} Z_{j n}^{\mathbf{0}} Z^\dagger_n}{2E_j(e^{E_j} - 1)}
	\bigg]
\\
	+ \sum_{m\neq n} 
	\frac{Z_{n}^{\mathbf{0}} Z^\dagger_m}{4E_n E_m}
	\frac{e^{-E_n t -\frac{\D_{mn}}{2}} - e^{-E_m t -\frac{\D_{nm}}{2}}}
		{e^{\frac{\D_{mn}}{2}} -e^{\frac{\D_{nm}}{2}}}
	J_{nm} 
	\bigg\} \, .
\label{eq:3ptansatz}
\end{multline}
In order to make the expressions more manageable, we introduce the following substitutions
\begin{align}
z_n &\equiv \frac{Z_n}{\sqrt{2E_n}},\\
g_{nm} &\equiv \frac{J_{nm}}{\sqrt{4E_nE_m}},\\
d_{n} &\equiv  Z_nZ_{J:n}^\dagger + Z_{J:n} Z_n^\dagger + Z_n Z_n^\dagger \langle \Omega | \mathcal{J} | \Omega \rangle \nonumber\\&\phantom{=}
	+\sum_{j} \frac{Z_n Z^\dagger_{nj} J_j^\dagger + J_j Z_{jn} Z_n^\dagger}{2E_j(e^{E_j}-1)}\, ,
\end{align}
leading to a two-point functional form
\begin{equation}
C(t,\mathbf{0}) = \sum_n z_n^\mathbf{0} z_n^\dagger e^{-E_nt}\, ,
\label{eq:2ptredefine}
\end{equation}
and the resulting numerator expression is
\begin{multline}\label{eq:3ptansatz_short}
	N(t)=\sum_{n} \left[(t-1) z_n g_{nn} z^\dagger_n + d_n\right]e^{-E_nt}
\\
	+\sum_{n\neq m} z_n g_{nm} z_m^\dagger 
		\frac{e^{-E_n t}e^{\frac{\Delta_{nm}}{2}} - e^{-E_mt}e^{\frac{\Delta_{mn}}{2}}}
			{e^\frac{\Delta_{mn}}{2} -e^\frac{\Delta_{nm}}{2}}.
\end{multline}

The simplicity of our method is recovered by substituting Eqs.~(\ref{eq:2ptredefine}) and (\ref{eq:3ptansatz_short}) into Eq.~(\ref{eq:dm_dl}).
All the contributions from regions II, III and IV are encoded in the $d_n$ constants.
These contributions, as well as the other excited-state contributions, are suppressed in the difference in the two terms in Eq.~\eqref{eq:dm_dl}.
It is straightforward to show that in the long-time limit, we recover the ground-state matrix element of interest
\begin{align}
\lim_{t\rightarrow\infty}\frac{\partial m^{\it eff}_\lambda(t,\tau)}{\partial \lambda}\bigg|_{\l=0} 
	= g_{00}
	= \frac{J_{00}}{2E_0}.
\end{align}

\subsection{Implementation \label{sec:implement}}

The numerical implementation of this  method is straightforward. What is needed is the construction of the derivative correlation function, Eq.~\eqref{eq:dc_dl_noVac}. We provide an explicit example of a nonscalar current $J_\G$ with interpolating operators coupling to the proton. Standard proton creation and annihilation operators are given by
\begin{align}
\bar{\mathcal{N}}_{\gp}(x) &= \epsilon_{\ip\jp\kp} 
	\left(\bar{u}^{\ip}_{\ap}(x) \G^{src}_{\ap\bp} \bar{d}^\jp_\bp(x) \right) P^{src}_{\gp\rp} \bar{u}^\kp_\rp(x)\, ,
\nonumber\\
\mathcal{N}_{\g}(y) &= -\epsilon_{ijk} 
	\left(u^i_\a(y) \G^{snk}_{\a\b} d^j_\b(y) \right) P^{snk}_{\g\rho} u^k_\rho(y)\, ,
\end{align}
where $u(x)$ and $d(x)$ are up and down quark field operators at $x$.
$\G^{src}$, $\G^{snk}$, $P^{src}$ and $P^{snk}$ are spin projectors used to project onto the total spin of the proton. For example, working in the Dirac basis, the dominant spin-up local interpolating field can be constructed using~\cite{Basak:2005aq,Basak:2005ir} (where the spinor indices are labeled 1,2,3,4)
\begin{align}
&P^{src} = \delta_{\mu,1}\, ,&
&\G^{src} = \begin{pmatrix}
0& 1& 0& 0\\
-1& 0& 0& 0\\
0& 0& 0& 0\\
0& 0& 0& 0
\end{pmatrix}\, ,&
\end{align}
and similar operators for the sink. Denoting the up and down quark propagators as
\begin{align}
U(y,x)^{i\ip}_{\a\ap} &= \bcontraction{}{u}{^i_\a(y)}{\bar{u}} u^i_\a(y) \bar{u}^{\ip}_{\ap}(x)\, ,
\nonumber\\
D(y,x)^{i\ip}_{\a\ap} &= \bcontraction{}{d}{^i_\a(y)}{\bar{d}} d^i_\a(y) \bar{d}^{\ip}_{\ap}(x)\, ,
\end{align}
the proton correlation function is  given by
\begin{multline}
C_{\g\gp} = \epsilon_{ijk} P^{snk}_{\g\rho}  \Big[
	(\G^{snk} D)^{i\ip}_{\a\bp} (U \G^{src})^{j\jp}_{\a\bp} U^{k\kp}_{\rho\rp}
\\
	+ (U\G^{src})^{k\kp}_{\rho\bp} U^{i\ip}_{\a\rp} (\G^{snk} D)^{j\jp}_{\a\bp}
	\Big] P^{src}_{\gp\rp} \epsilon_{\ip\jp\kp}  \, .
\end{multline}
The derivative correlation function ($-\partial_\l C |_{\l=0}$) is trivially determined. Applying the partial derivative in Eq.~\eqref{eq:dc_dl} at the level of the path integral, one immediately observes that the derivative correlation function is obtained with the replacement of one of the quark propagators in the two-point correlation function with a \textit{Feynman-Hellmann} (FH) propagator, summed over all possible insertions, where the FH propagator is simply a sequential propagator which is also summed over the current insertion time
\begin{equation}
S^\G(y,x) = \sum_{z=(t_z,\mathbf{z})} S(y,z) \G S(z,x)\, .
\end{equation}
This extra sum leads to an $\mc{O}(t)$ stochastic enhancement of the resulting derivative correlation function as compared to the standard method with a sequential propagator. The idea of using this propagator in a two-point function can be traced back to Ref.~\cite{Maiani:1987by}, where the equivalent of Eq.~\eqref{eq:R(t)} is approximated with its long-time limit and computed for various hadronic matrix elements. 
The idea was further discussed in Ref.~\cite{Bulava:2011yz} and applied to $B^*B\pi$ couplings in Ref.~\cite{Bernardoni:2014kla}.
This idea has been extended recently in Refs.~\cite{deDivitiis:2012vs,Alexandrou:2013xon} with an application to derivatives with respect to $Q^2$ of mesonic structure functions in Ref.~\cite{deDivitiis:2012vs}.

For our present example, the proton derivative correlation function with a current insertion on the down quark is given by $D\rightarrow D^\G$
\begin{multline}
C_{\g\gp}^{\G_d} = \epsilon_{ijk} P^{snk}_{\g\rho}  \Big[
	(\G^{snk} D^\G)^{i\ip}_{\a\bp} (U \G^{src})^{j\jp}_{\a\bp} U^{k\kp}_{\rho\rp}
\\
	+ (U\G^{src})^{k\kp}_{\rho\bp} U^{i\ip}_{\a\rp} (\G^{snk} D^\G)^{j\jp}_{\a\bp}
	\Big] P^{src}_{\gp\rp} \epsilon_{\ip\jp\kp}  \, ,
\end{multline}
while for a current insertion on the up quark, one has
\begin{multline}
C_{\g\gp}^{\G_u} = \epsilon_{ijk} P^{snk}_{\g\rho}  \Big[
	(\G^{snk} D)^{i\ip}_{\a\bp} (U^\G \G^{src})^{j\jp}_{\a\bp} U^{k\kp}_{\rho\rp}
\\
	+(\G^{snk} D)^{i\ip}_{\a\bp} (U \G^{src})^{j\jp}_{\a\bp} U^{\G,k\kp}_{\rho\rp}
\\
	+ (U^\G \G^{src})^{k\kp}_{\rho\bp} U^{i\ip}_{\a\rp} (\G^{snk} D)^{j\jp}_{\a\bp}
\\
	+ (U\G^{src})^{k\kp}_{\rho\bp} U^{\G,i\ip}_{\a\rp} (\G^{snk} D)^{j\jp}_{\a\bp}
	\Big] P^{src}_{\gp\rp} \epsilon_{\ip\jp\kp}  \, .
\end{multline}
These correlators are functions of the source-sink separation time in contrast to the fixed source-sink separation time dependence of the standard three-point correlators constructed with sequential propagators. It is trivial to generalize this construction to an arbitrary correlation function with the successive replacement of each quark propagator with its respective FH propagator.

For many matrix elements, one must also consider contributions from disconnected diagrams. While disconnected diagrams are stochastically noisier, we may be able to improve upon the general method as we have the freedom to compute the disconnected quark loop as a function of Euclidean time. Instead of summing over all time as in Eq.~\eqref{eq:dc_dl_noVac}, we can explicitly choose to only sum over the time window $0 < t^\prime < t$, thus including only the contributions of interest [Eq.~\eqref{eq:NI}]. It is worth exploring if this suggestion provides an improved determination of the disconnected contributions.

\section{An application to the nucleon axial charge \label{Sec:3}}

We demonstrate the use of this method by computing the nucleon isovector axial charge on one of the publicly available 2+1+1 HISQ~\cite{Follana:2006rc} ensembles from the MILC Collaboration~\cite{Bazavov:2012xda}, with $a\simeq0.15$~fm, $m_\pi\simeq310$~MeV and a lattice volume of $16^3\times48$.
The present work utilizes 1960 configurations with six sources per configuration.
The HISQ configurations are first gradient flowed to smear out the UV fluctuations.
M\"{o}bius domain-wall fermion (MDWF)~\cite{Brower:2012vk} propagators are then solved with the QUDA library~\cite{Clark:2009wm} with multi-GPU support~\cite{Babich:2011np}.
The action and efficiency of the software was described in Ref.~\cite{Berkowitz:2017opd}.
The valence quark mass is tuned such that the MDWF pion mass matches the taste-5 HISQ pion mass within 1\%.
We then construct the isovector nucleon two-point and derivative FH correlation functions.
We further double our statistics by including the time-reversed correlation functions constructed with negative-parity nucleon operators.
The calculation requires solving both the regular and the FH propagator for each source.
The motivation and advantages of such a mixed action were described in Ref.~\cite{Berkowitz:2017opd}.
This action has also been used to compute the $\pi^-\rightarrow\pi^+$ matrix element relevant to $0\nu\b\b$ (neutrinoless double beta-decay)~\cite{Nicholson:2016byl}.

\subsection{Fit strategy}
We extract the isovector nucleon charge by applying two analysis strategies:
we apply a standard frequentist analysis of our results as well as a Bayesian constrained fit with Gaussian priors on the two- and three-point correlation functions following the framework of Ref.~\cite{Lepage:2001ym}. While we find consistent answers with both analysis strategies, the Bayesian analysis allows us to more rigorously and completely explore the fitting systematics, thus demonstrating the efficacy of this new method. Controlling fitting systematics for nucleons is critical for current and future LQCD calculations. Similar ideas of Bayesian constrained curve fits were explored in Ref.~\cite{Chen:2004gp} and recently used in Ref.~\cite{Yoon:2016jzj}.

In our present Bayesian analysis, we observe stability in all ground-state parameters under variations of the number of time slices in our data, and number of states in our fit \textit{Ansatz}, demonstrating complete control over systematic uncertainty originating from the fitting procedure.  We rate the quality of our fits by the $Q$-statistic, which is related to the frequentist $p$-value as defined in Eq.~(B4) of Ref.~\cite{Bazavov:2016nty} and consider only fits with $Q > 0.1$. We select results from different fit \textit{Ans\"atze} by comparing Bayes factors, and make distinctions between models if the Bayes factor is larger than 3.  Constrained curve fits are implemented by the software package  {\tt lsqfit}~\cite{lsqfit}.

In the following sections, the fit procedure is first discussed for the two-point correlation function. The preferred two-point fit is then performed simultaneously with the three-point correlator leading to our final value for $g_A$. We adopt the notation of placing a tilde on top of the parameter (e.g. $E_0$) to denote its prior distribution (e.g. $\tilde{E}_0$), and a hat (e.g. $\hat{E}_0$) to denote its posterior distribution.

\subsection{Two-point correlator analysis}
\label{sec:2ptcorr}
\subsubsection{Fit strategy}
The two-point fit \textit{Ansatz} is given by Eq.~(\ref{eq:2ptredefine}) and has a sum over the infinite tower of states. 
We focus on the case of zero momentum insertion and drop the superscript momentum label, $z_n^\mathbf{0}\rightarrow z_n$. In the following section, we detail how the priors are set for the infinite number of states, and quantitatively show where the series may be truncated.

\begin{figure}[h]
	\centering
		\includegraphics[width=1.0\columnwidth]{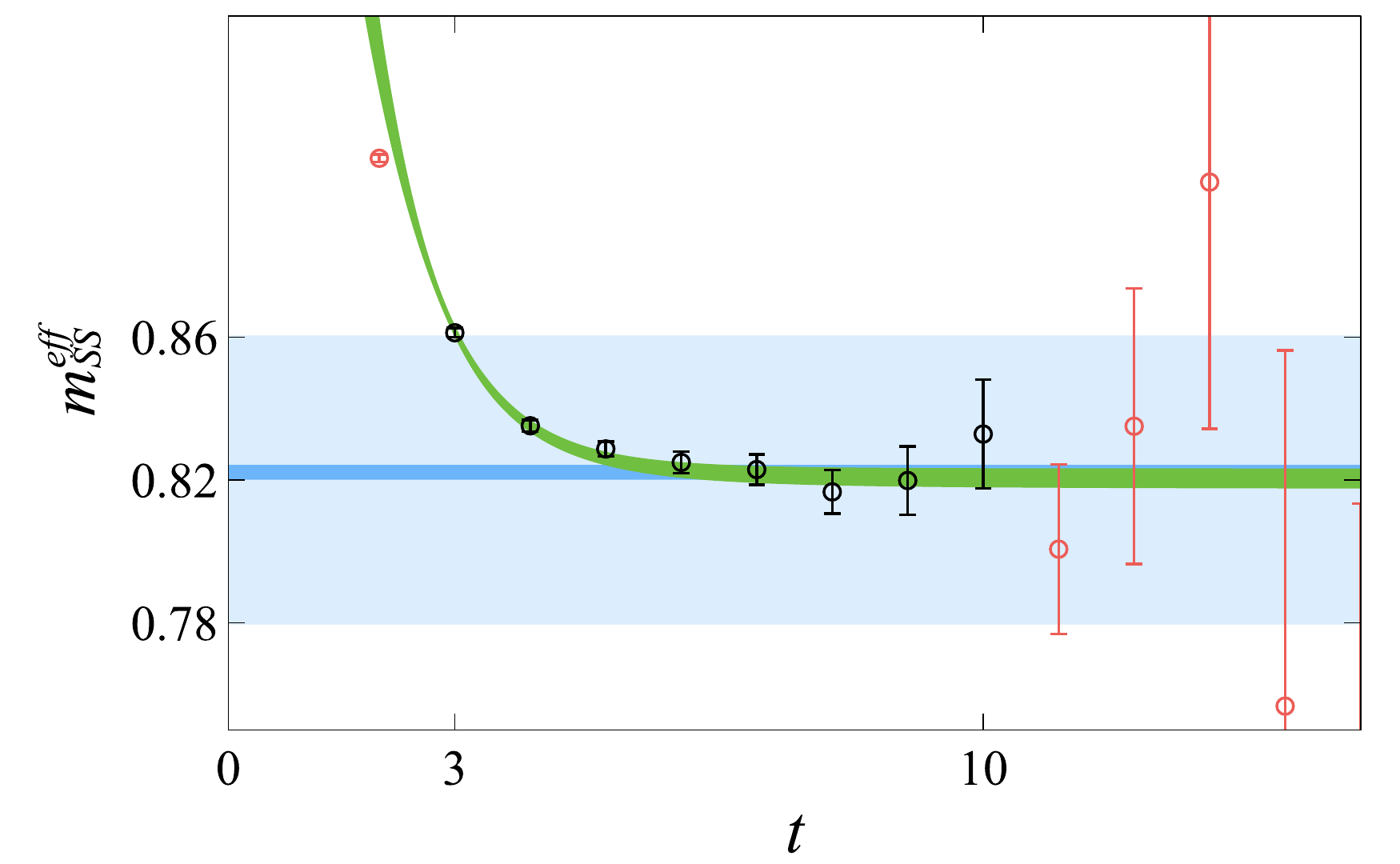}
	\caption{Plot of the smeared-sink effective mass as a function of source-sink separation time $t$. The black points highlight the data used in the fit presented. The light blue bands indicate the $1\sigma$ width of the ground-state energy prior $\tilde{E}_0$, the dark blue bands show the central value and $1\sigma$ uncertainty of the corresponding posterior distribution. The green bands are the resulting fit curves from a simultaneous fit to the smeared- and point-sink two-point correlation functions.}
	\label{fig:meff}
\end{figure}

\begin{figure}[h]
	\centering
		\includegraphics[width=1.0\columnwidth]{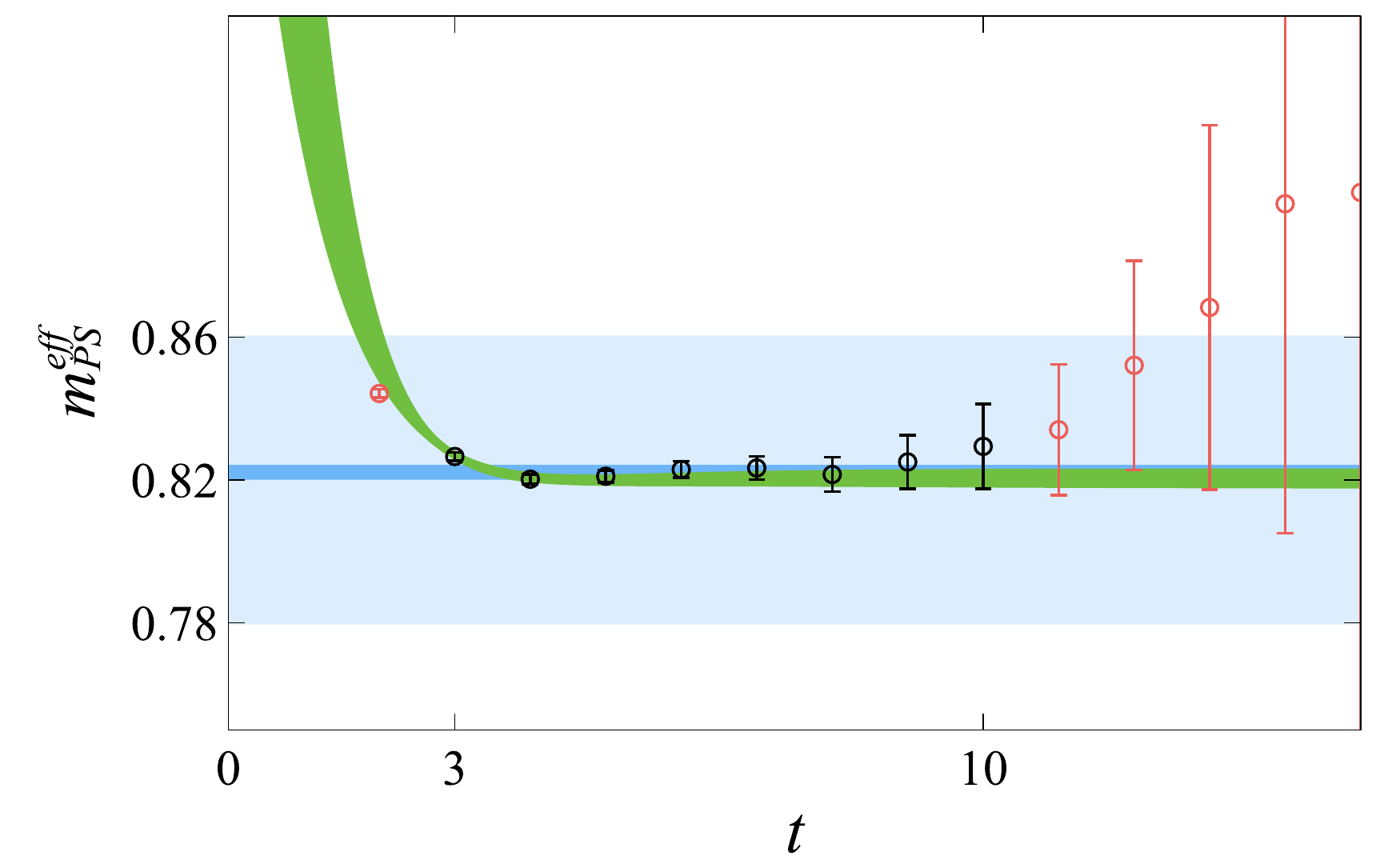}
	\caption{Same as Fig.~\ref{fig:meff} but for the point-sink effective mass.}
	\label{fig:meff_ps}
\end{figure}

We begin our two-point analysis by looking at the effective mass as defined in Eq.~(\ref{eq:meff}) and shown in Figs.~\ref{fig:meff} and \ref{fig:meff_ps}. We observe a plateau indicating that the ground-state energy has a value of approximately $E_0\approx0.82$. We set a loose prior of $\tilde{E}_0 = 0.82(4)$ as indicated by the light blue band. The width of the prior is observed to be approximately 1 order of magnitude larger than the resulting posterior distribution, which is indicated by the dark blue band.

\begin{figure}[h]
	\centering
		\includegraphics[width=1.0\columnwidth]{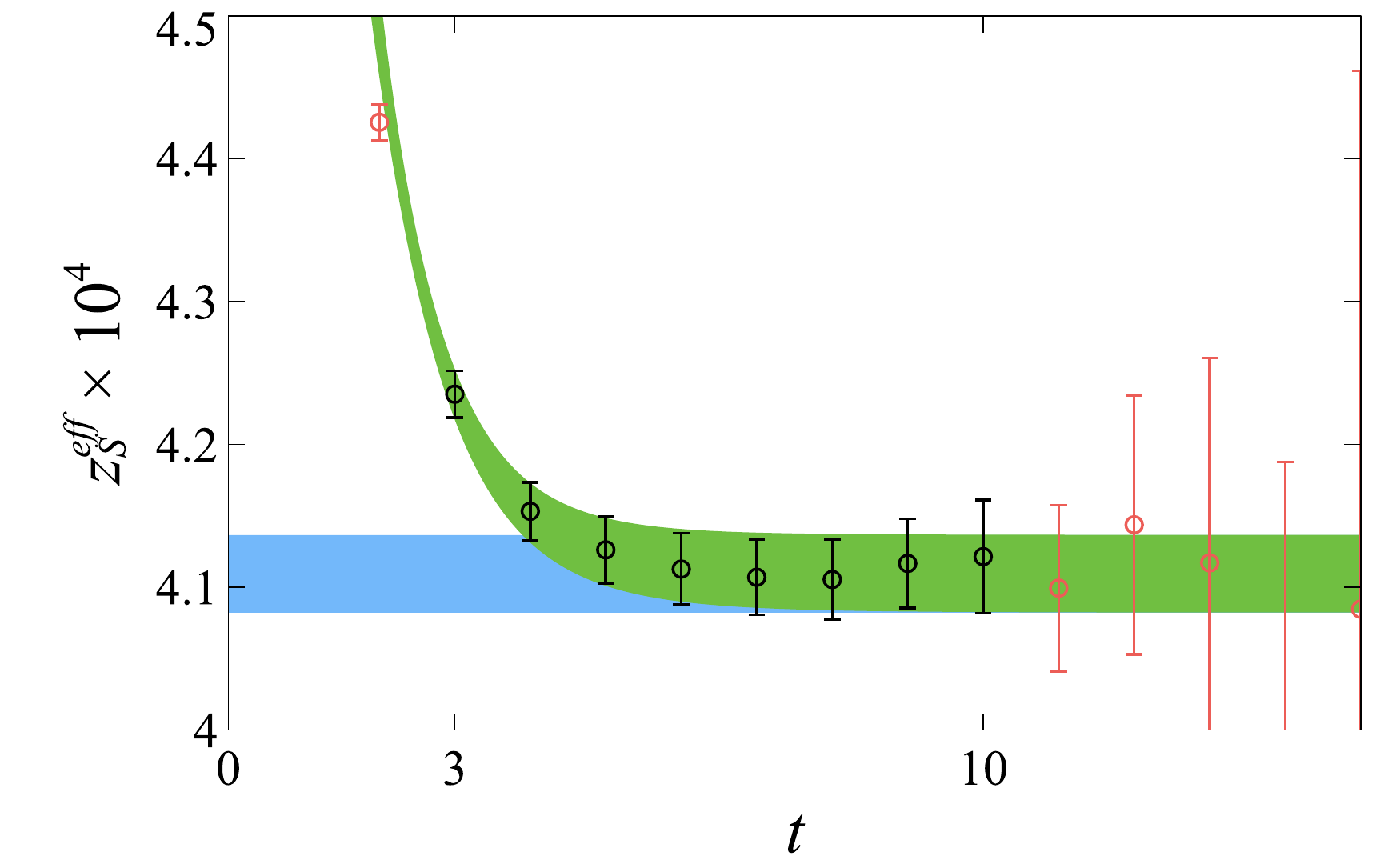}
	\caption{Plot of the scaled two-point correlation function as a function of source-sink separation time $t$ for the smeared overlap factor. The color scheme follows Fig.~\ref{fig:meff}. The prior width exceeds the range of the $y$ axis, and is therefore not included in the plot.}
	\label{fig:zeff}
\end{figure}

\begin{figure}[h]
	\centering
		\includegraphics[width=1.0\columnwidth]{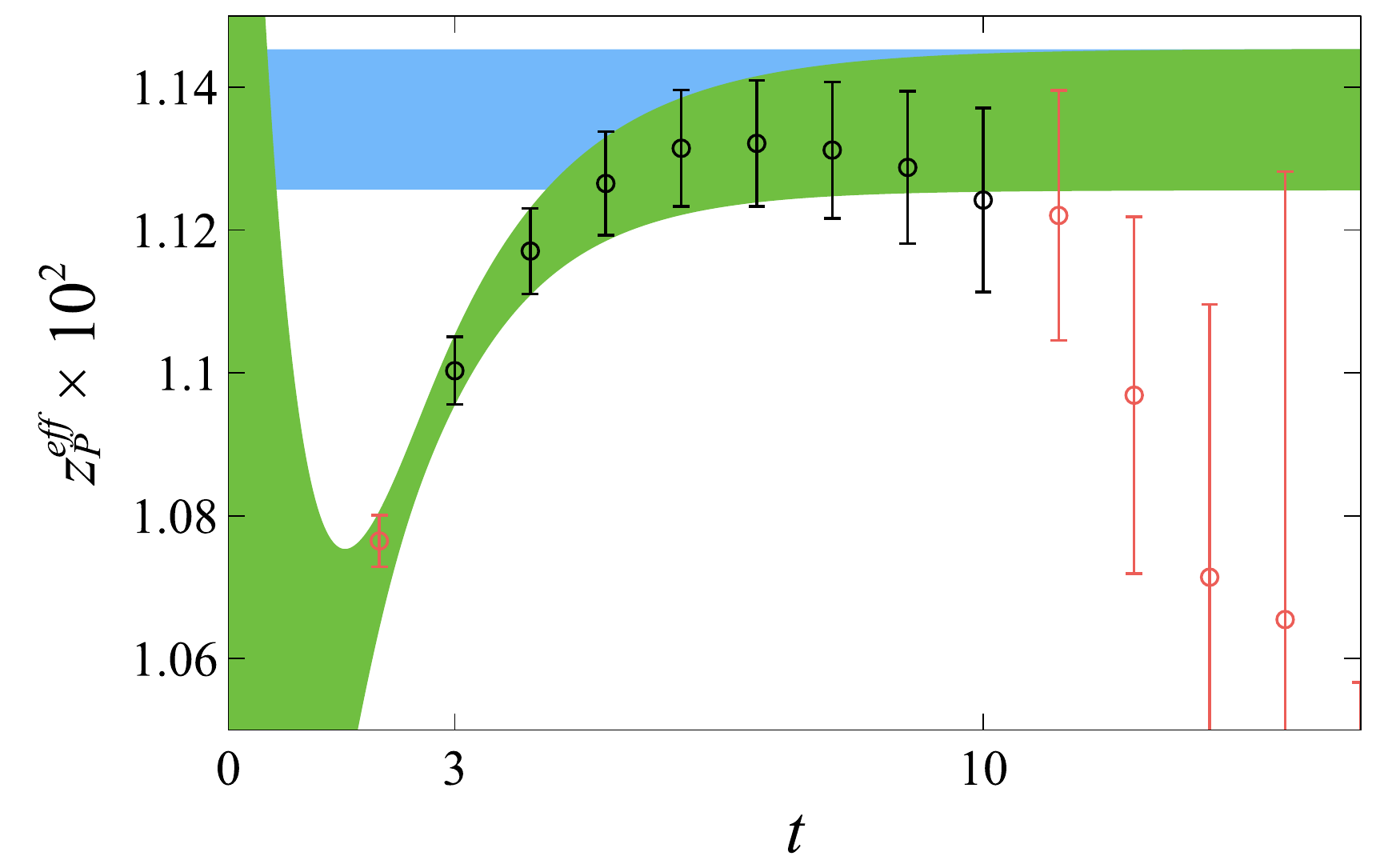}
	\caption{Same as Fig.~\ref{fig:zeff} but for the point-like overlap factor.}
	\label{fig:zeff_p}
\end{figure}

Using the estimate of $E_0$, we construct the scaled two-point function in order to set prior distributions for $\tilde{z}_0^S$ and $\tilde{z}_0^P$, which are respectively the ground-state smeared and point overlap factors. We define the scaled correlators as
\begin{align}
z^{\it eff}_S(t) = & \sqrt{C_{SS}(t) e^{\tilde{E_0}t}},\\
z^{\it eff}_P(t) = & C_{PS}(t) e^{\tilde{E_0}t}/z^{\it eff}_S(t),
\end{align}
where $C_{SS}(t)$ and $C_{PS}(t)$ are the smeared-smeared and point-smeared two-point functions, respectively. From the redefinition of Eq.~(\ref{eq:2ptredefine}), the scaled correlators plateau to $z_0^{S(P)}$ in the $t\rightarrow \infty$ limit. From Figs.~\ref{fig:zeff}~and~\ref{fig:zeff_p}, we assign approximate values for the overlap factors, and set prior widths to half the magnitude of the expected central value leading to $\tilde{z}_0^S= 0.0004(2)$ and $\tilde{z}_0^P=0.010(5)$. These priors are unconstraining since the widths of the posterior distributions are approximately 2 orders of magnitude smaller.

We introduce priors for the excited-state energies as energy splittings, enforcing a hierarchy of states. This is achieved by setting the energy splitting with a lognormal prior distribution. For the nucleon on a lattice with a $\approx 310$~MeV pion mass and a box size of $\approx 2.4$~fm, we prior the first three excited states in the following order: the Roper resonance ($\approx 450~\text{MeV}$ from the ground state), the two-pion excitation ($\approx 180~\text{MeV}$ from the Roper), and the $L=1$ one-pion excitation ($\approx 110~\text{MeV}$ from the two-pion excitation), with prior widths which accommodate a one-pion splitting (310~MeV) within $1\sigma$. For the fourth excited state onward, the prior for the energy splitting is set to the one-pion splitting while accommodating the two-pion splitting within $1\sigma$.

The excited-state overlap factors can enter with an unknown sign (point-smeared); therefore, so as to not bias our expectation, we set the prior central values of the excited-state overlap factors to zero. Due to smearing, we expect less overlap with excited states, and therefore set the width of $\tilde{z}_n^S$ to half the value of the $\tilde{z}_0^S$ central value. For the point-like overlap factor, we expect equal support from all states and therefore set the width of $\tilde{z}_n^P$ to the value of the $\tilde{z}_0^P$ central value.

\subsubsection{Discussion}
\begin{figure}[h]
	\centering
		\includegraphics[width=1.0\columnwidth]{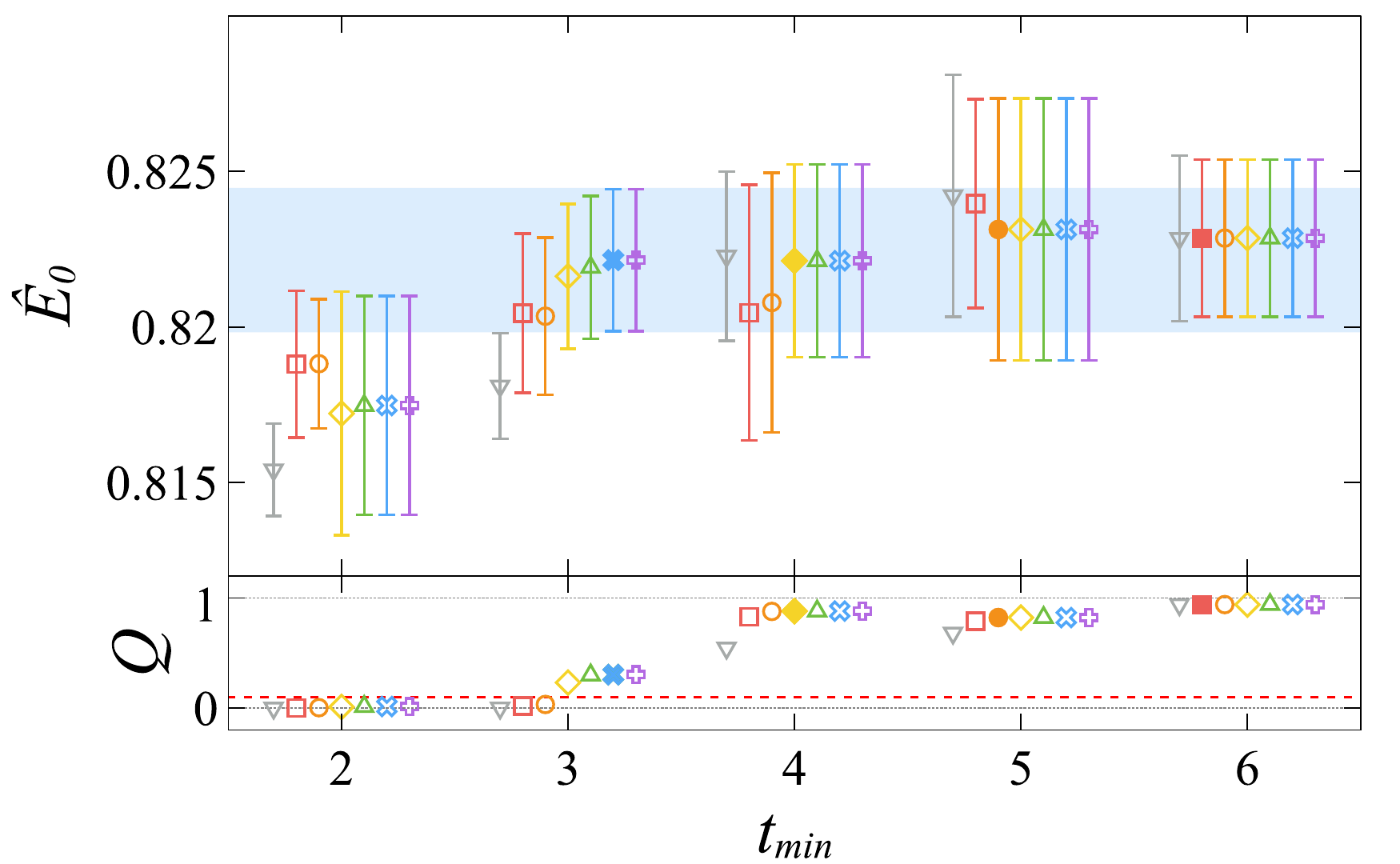}
	\caption{Stability plots for the two-point correlation function. Stability plot under varying $t_{min}$ with fixed $t_{max}=10$ is shown for fits with two states ({\color{kngrey}\ding{116}}), three states ({\color{knred} \ding{110}}), four states ({\color{knorange} \ding{108}}), five states ({\color{knyellow} \ding{117}}), six states ({\color{kngreen} \ding{115}}), seven states ({\color{knblue} \ding{54}}), and eight states ({\color{knpurple}\ding{58}}). The corresponding $Q$-values of the fits are plotted below with the dashed line set at $Q=0.1$. The solid symbols mark the fit with the largest Bayes factor at fixed $t_{min}$. The blue band highlights the preferred fit with seven states at $t_{min}=3$ and guides the eye for observing stability.}
	\label{fig:2pt_stability}
\end{figure}

\begin{figure}[h]
	\centering
		\includegraphics[width=1.0\columnwidth]{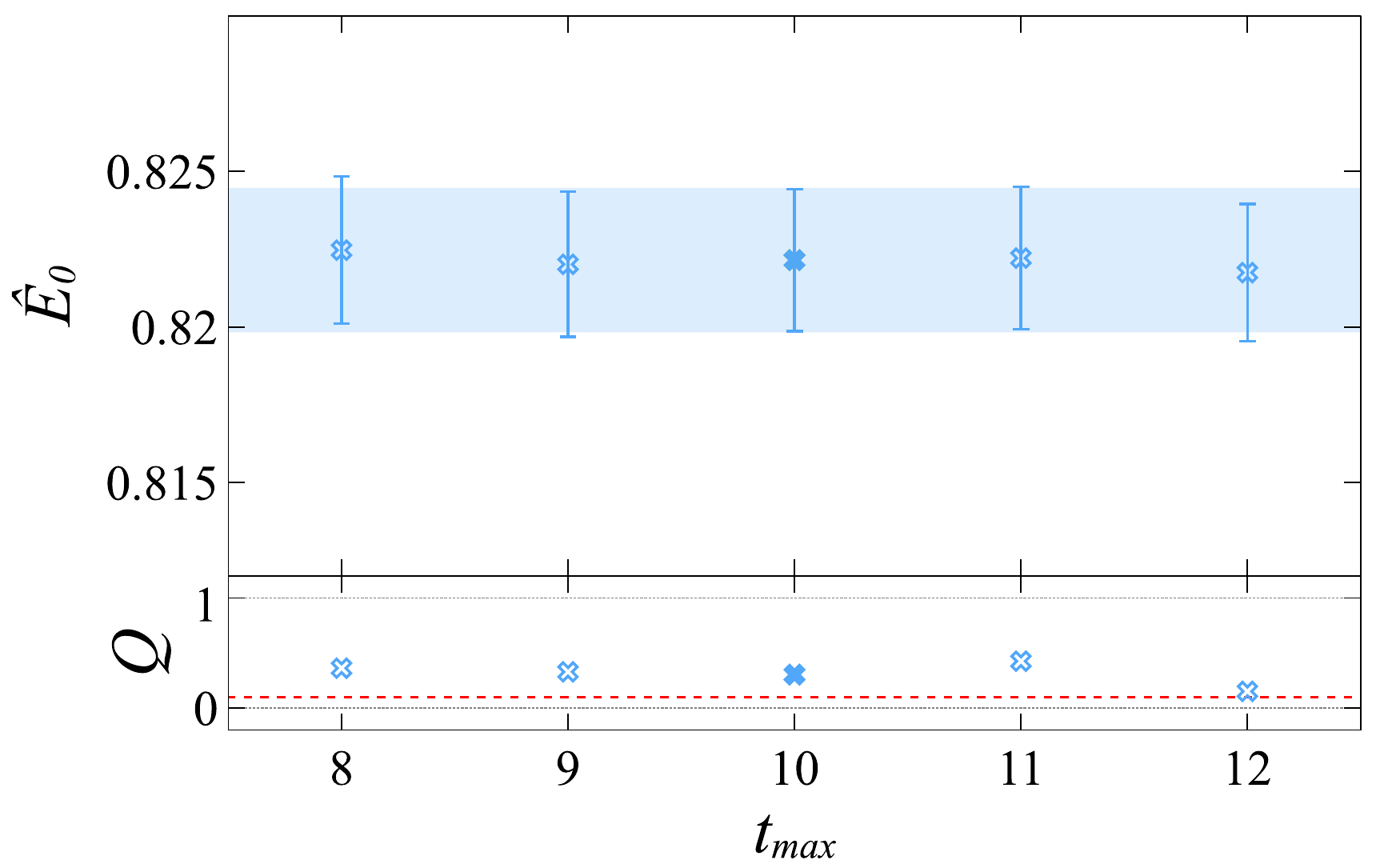}
	\caption{Stability plot under varying $t_{max}$ for the seven-state fit with $t_{min}=3$. The color scheme follows Fig.~\ref{fig:2pt_stability}. The solid symbol denotes the preferred two-point correlator fit.}
	\label{fig:2pt_stability_tmax}
\end{figure}

We study the stability of $\hat{E}_0$ from the two-point correlator fits under variation of the fit region and fit \textit{Ansatz} as shown in Figs.~\ref{fig:2pt_stability} and \ref{fig:2pt_stability_tmax}.

Perfect stability under varying the number of states is demonstrated from fits in Fig.~\ref{fig:2pt_stability} with $t_{min}\geq 3$, suggesting that we have controlled all the excited-state systematic uncertainty. The fits show that more excited states are needed to achieve stability when including data with smaller $t_{min}$ due to larger excited-state contamination. We also observe that for all fit variations at a fixed $t_{min}$ the Bayes factor always prefers the stable fit result with the lowest number of excited states, as indicted by the solid symbols. We conclude that the increased uncertainty and variation of the central value from fits beyond $t_{min}=3$ are only due to omitting parts of the usable data set. Therefore the preferred two-point fit is a seven-state fit with $t=[3, 10]$. We note that all fits with $t_{min}=2$ do not pass the $Q$-value criterion and are not considered. Similarly, Fig.~\ref{fig:2pt_stability_tmax} shows stability under varying $t_{max}$ for the preferred fit and suggests that $t_{max}=10$ is adequate.

The preferred fit is plotted in Figs.~\ref{fig:meff}--\ref{fig:zeff_p} as a green band and agrees with the data in the selected fit region, which is indicated by the black data points. 
The ground-state parameters $E_0$, $z_0^S$ and $z_0^P$ are recovered in the $t\rightarrow \infty$ limit from the effective mass and scaled correlators by construction.
This is demonstrated by observing that the green fit band overlaps with the dark blue posterior band asymptotically at large $t$.

\subsection{Three-point correlator analysis}
\subsubsection{Fit strategy}
The three-point fit \textit{Ansatz} is given by Eq.~(\ref{eq:3ptansatz_short}). 
The three-point fit \textit{Ansatz} introduces additional parameters $g_{nm}$ and $d_n$ which are not constrained by the two-point correlator. In the following section, we detail how priors are chosen for the additional parameters, and study stability under variations of fit region and number of states.

\begin{figure}[h]
	\centering
		\includegraphics[width=1.0\columnwidth]{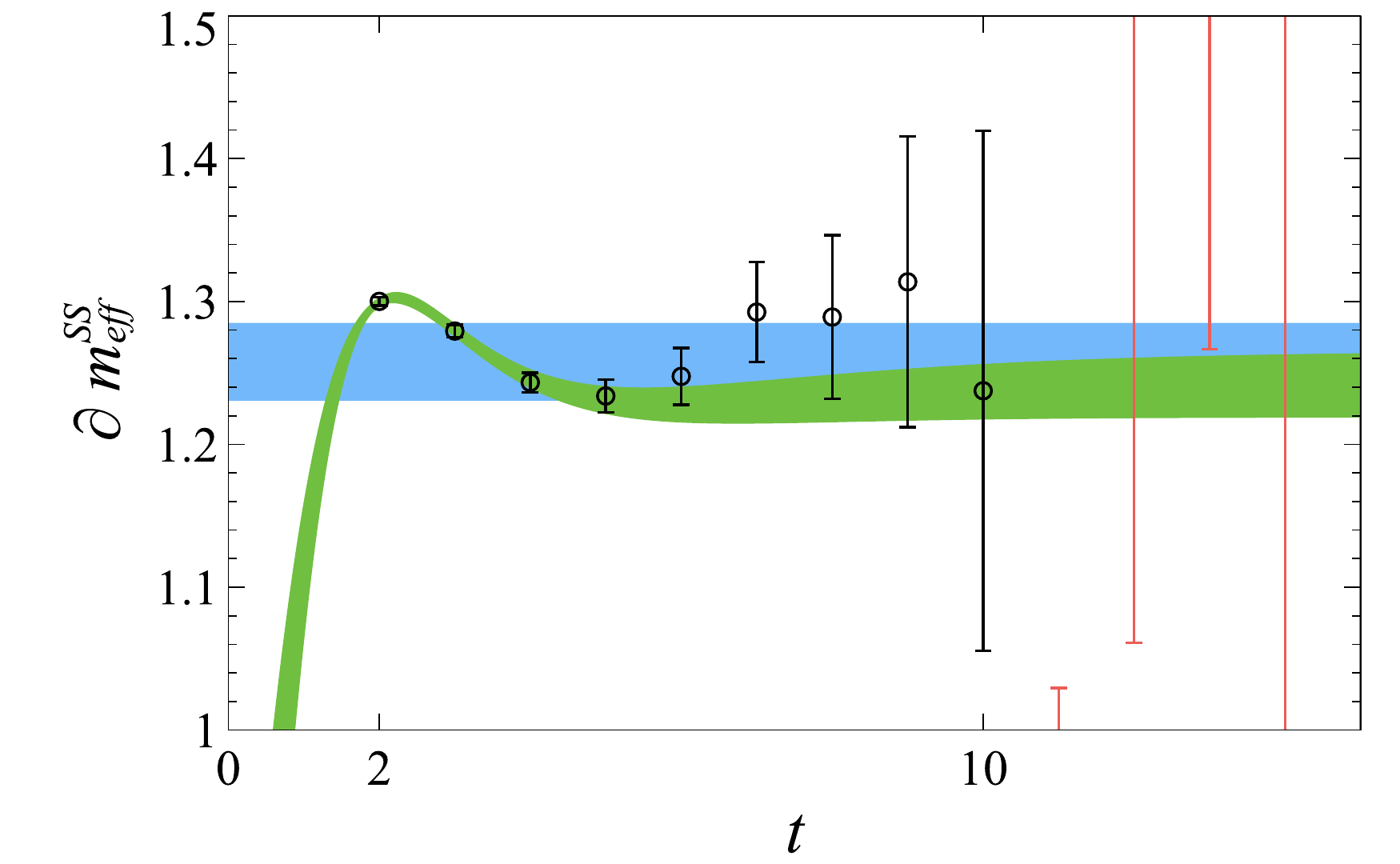}
	\caption{Plot of the derivative effective mass as a function of source-sink separation time $t$ for the smeared sink correlation functions. The color scheme follows Fig.~\ref{fig:meff}. The prior width exceeds the range of the $y$ axis, and is therefore not included in the plot.}
	\label{fig:dmeff}
\end{figure}

\begin{figure}[h]
	\centering
		\includegraphics[width=1.0\columnwidth]{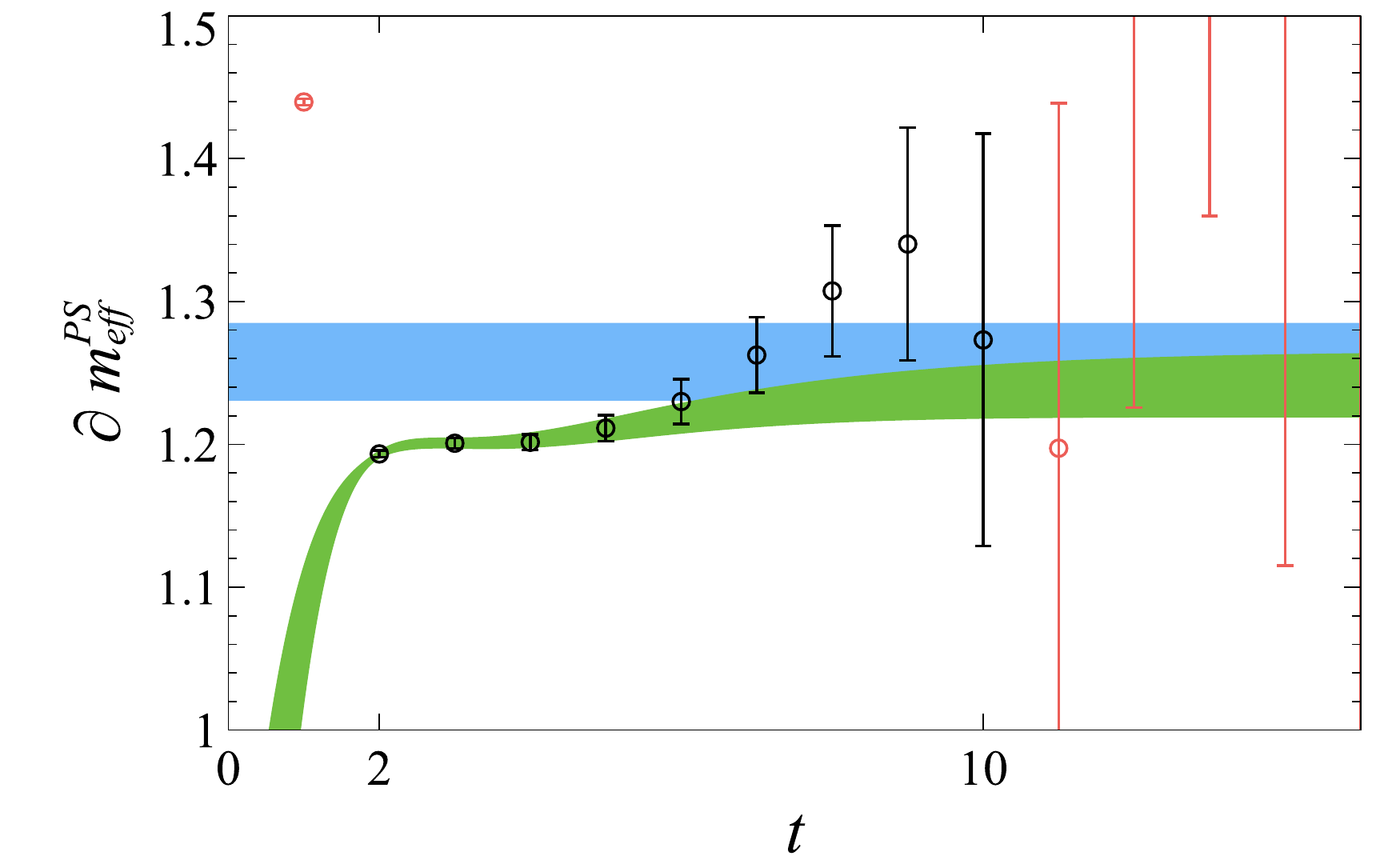}
	\caption{Same as Fig.~\ref{fig:dmeff} for the point-sink correlation function.}
	\label{fig:dmeff_ps}
\end{figure}

We set the ground-state prior $\tilde{g}_{00}$ by constructing the derivative effective mass given by Eq.~(\ref{eq:dmeff}). By construction the derivative effective mass plateaus to $g_{00}$ in the $t\rightarrow\infty$ limit. Motivated by Figs.~\ref{fig:dmeff} and \ref{fig:dmeff_ps}, we set $\tilde{g}_{00}=1.2(5)$ and observe that the prior widths are approximately 1 order of magnitude larger than the width of the posterior distributions.

\begin{figure}[h]
	\centering
		\includegraphics[width=1.0\columnwidth]{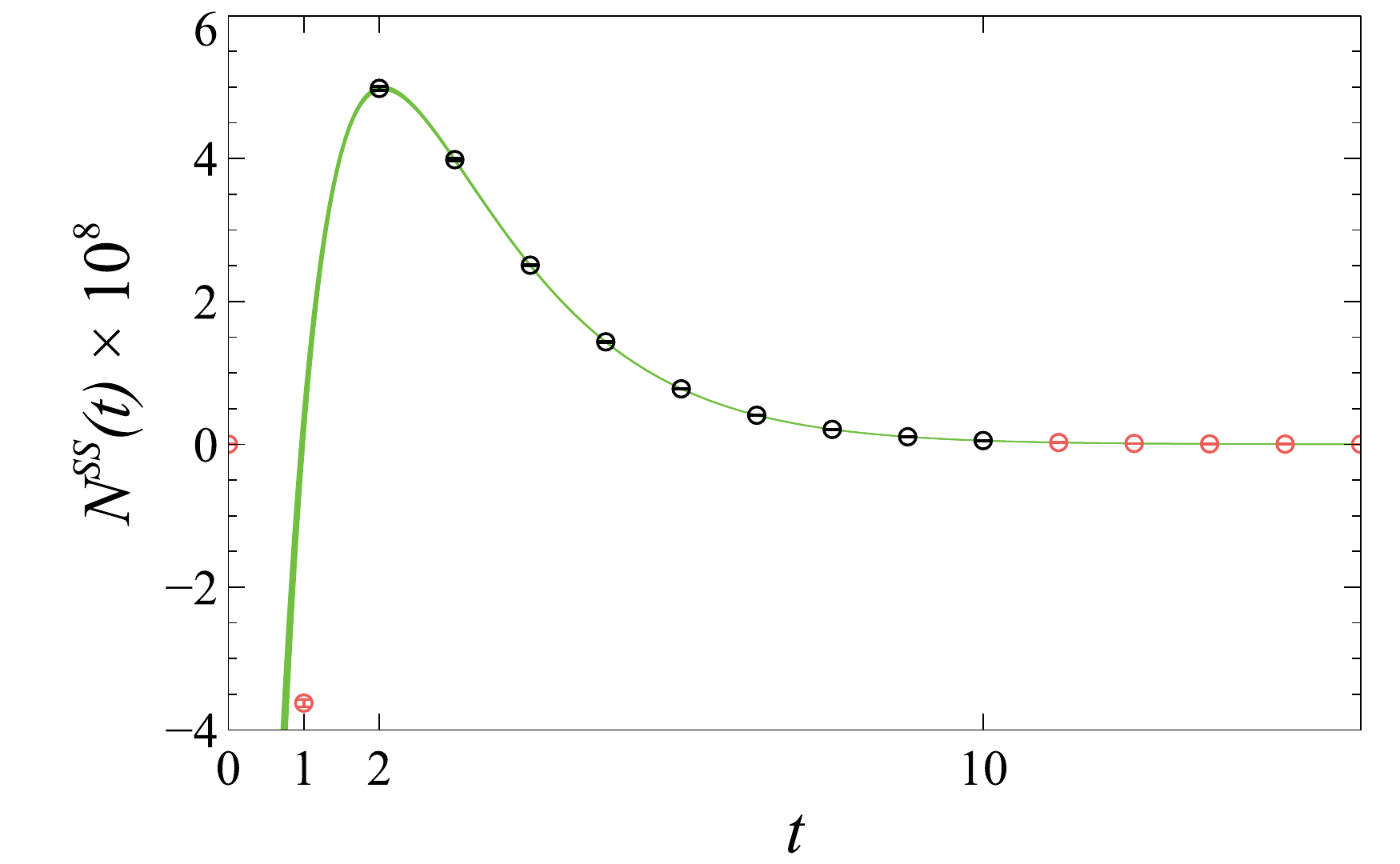}
	\caption{Plot of the three-point correlator as a function of source-sink separation time $t$ for the smeared-sink correlation functions. The color scheme follows Fig.~\ref{fig:meff}.
	Note the distinct behavior of the $N(t)$ correlation function at $t=1$ as discussed earlier.}
	\label{fig:N}
\end{figure}

\begin{figure}[h]
	\centering
		\includegraphics[width=1.0\columnwidth]{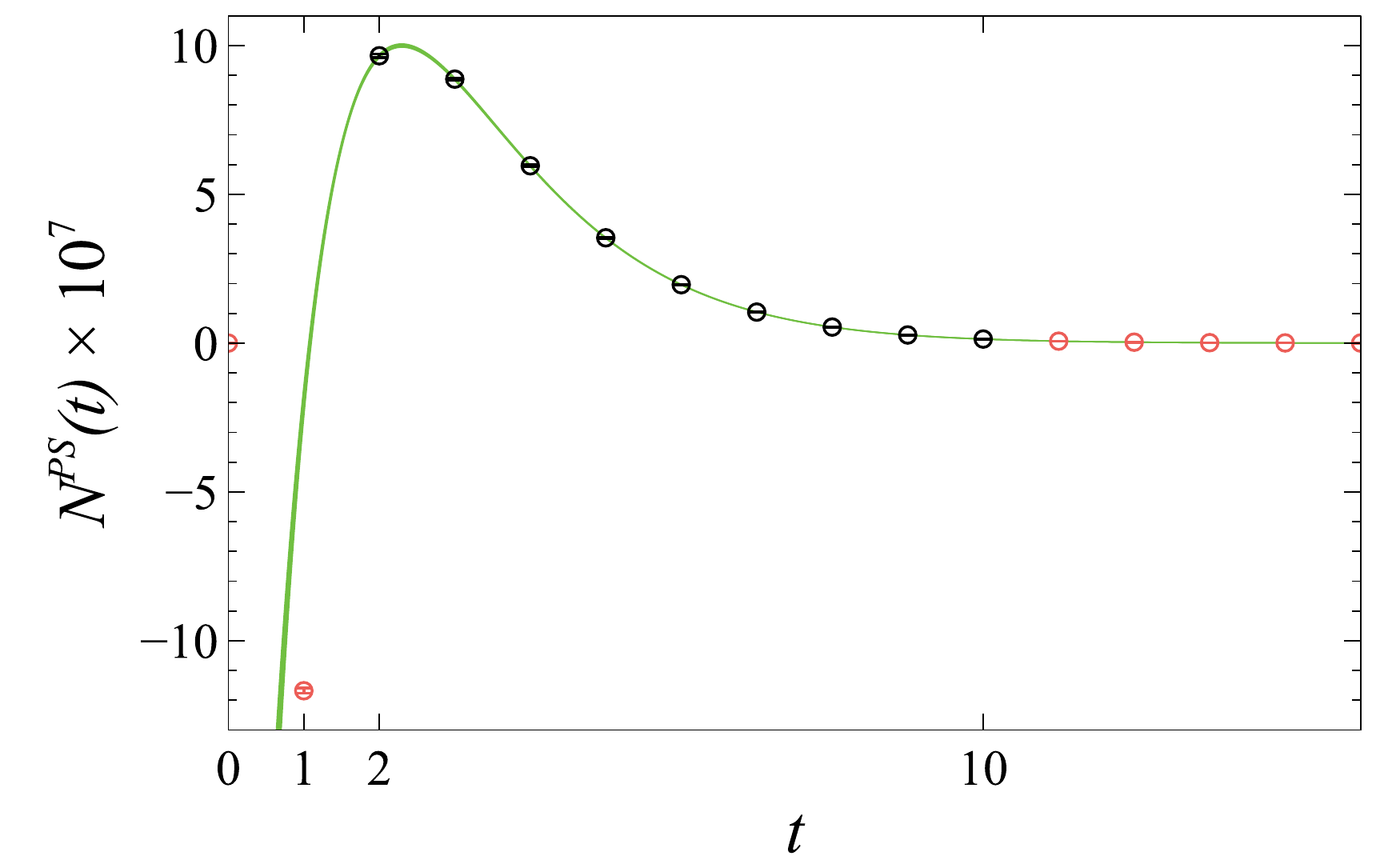}
	\caption{Same as Fig.~\ref{fig:N} for the point-sink $N(t)$  function.}
	\label{fig:N_ps}
\end{figure}

The prior for $d^{SS(PS)}_0$ is chosen by observing that
\begin{equation}
N^{SS(PS)}(t=1) =\sum_{n} d^{SS(PS)}_n e^{-E_n}
\label{eq:contactcoeff}
\end{equation}
and assuming that the contribution to the infinite sum is primarily from the $n=0$ state due to operator smearing. This unique feature of the correlator at $t=1$ is exemplified in Figs.~\ref{fig:N} and \ref{fig:N_ps}. Therefore we set the central value of $\tilde{d}^{SS(PS)}_0 = N^{SS(PS)}(1) e^{E_0} / 2$ with a prior width accommodating $d^{SS(PS)}_0=0$ within $1\sigma$. 

For the excited-state priors, we set $\tilde{g}_{nm} = 0(1)$ to be the same order of magnitude as $\tilde{g}_{00}$. The overlap factor $d^{SS(PS)}_n$ is set with a central value of $\tilde{d}^{SS(PS)}_n = 0$ and with the same width as $\tilde{d}^{SS(PS)}_0$. 
This choice follows the same logic used to set the priors for $z_n^S$.

\subsubsection{Discussion}
\begin{figure}
	\centering
		\includegraphics[width=1.0\columnwidth]{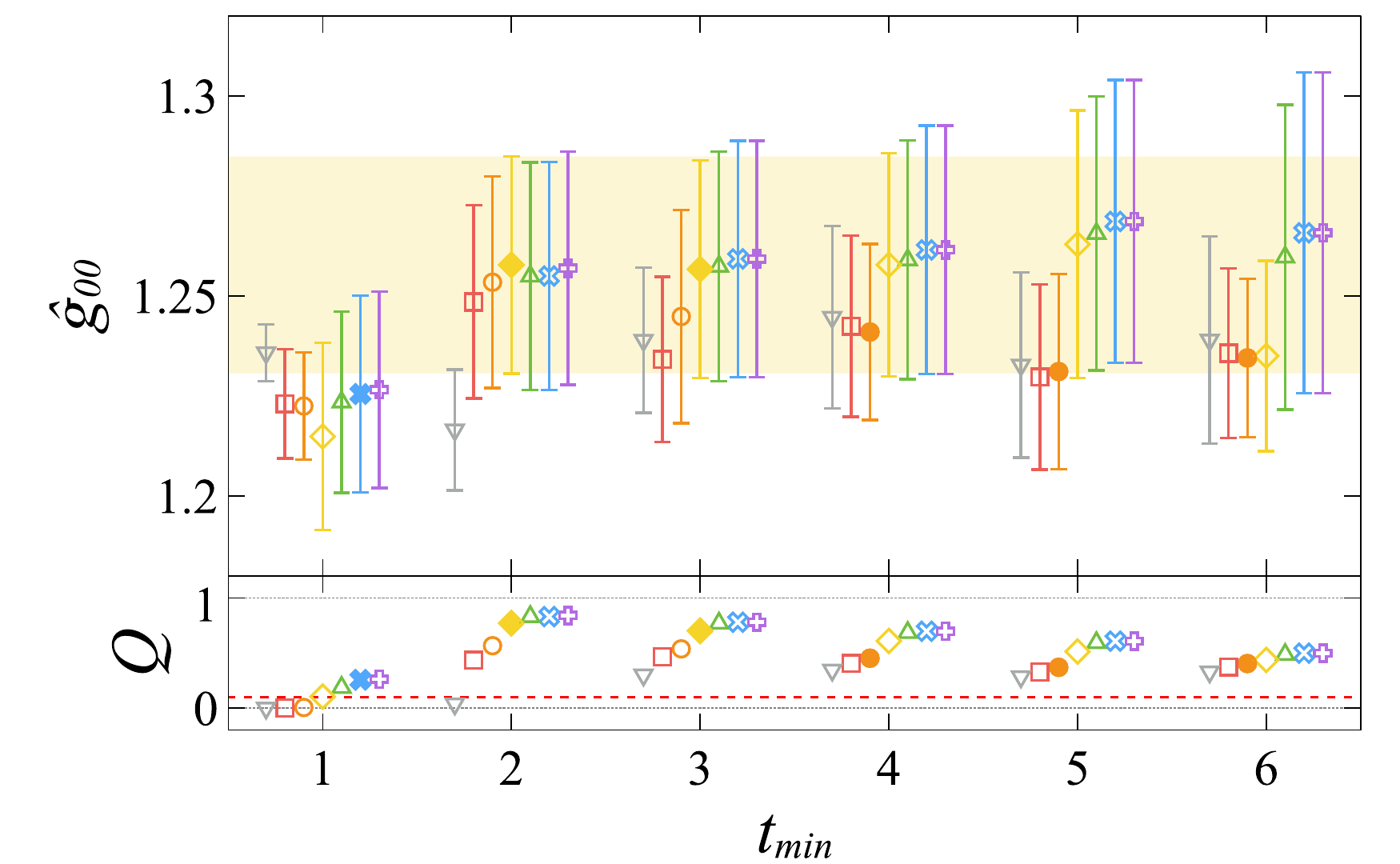}
	\caption{Stability plots for the three-point correlation function. Stability plot under varying $t_{min}$ with fixed $t_{max}=10$ is shown for fits with two to eight states analogous to Fig.~\ref{fig:2pt_stability}. The corresponding $Q$-values of the fits are plotted below with the dashed line set at $Q=0.1$. The solid symbols mark the fit with the largest Bayes factor at fixed $t_{min}$. The yellow band highlights the preferred fit with five states at $t_{min}=2$ and guides the eye for observing stability.}
	\label{fig:3pt_stability}
\end{figure}

\begin{figure}[h]
	\centering
		\includegraphics[width=1.0\columnwidth]{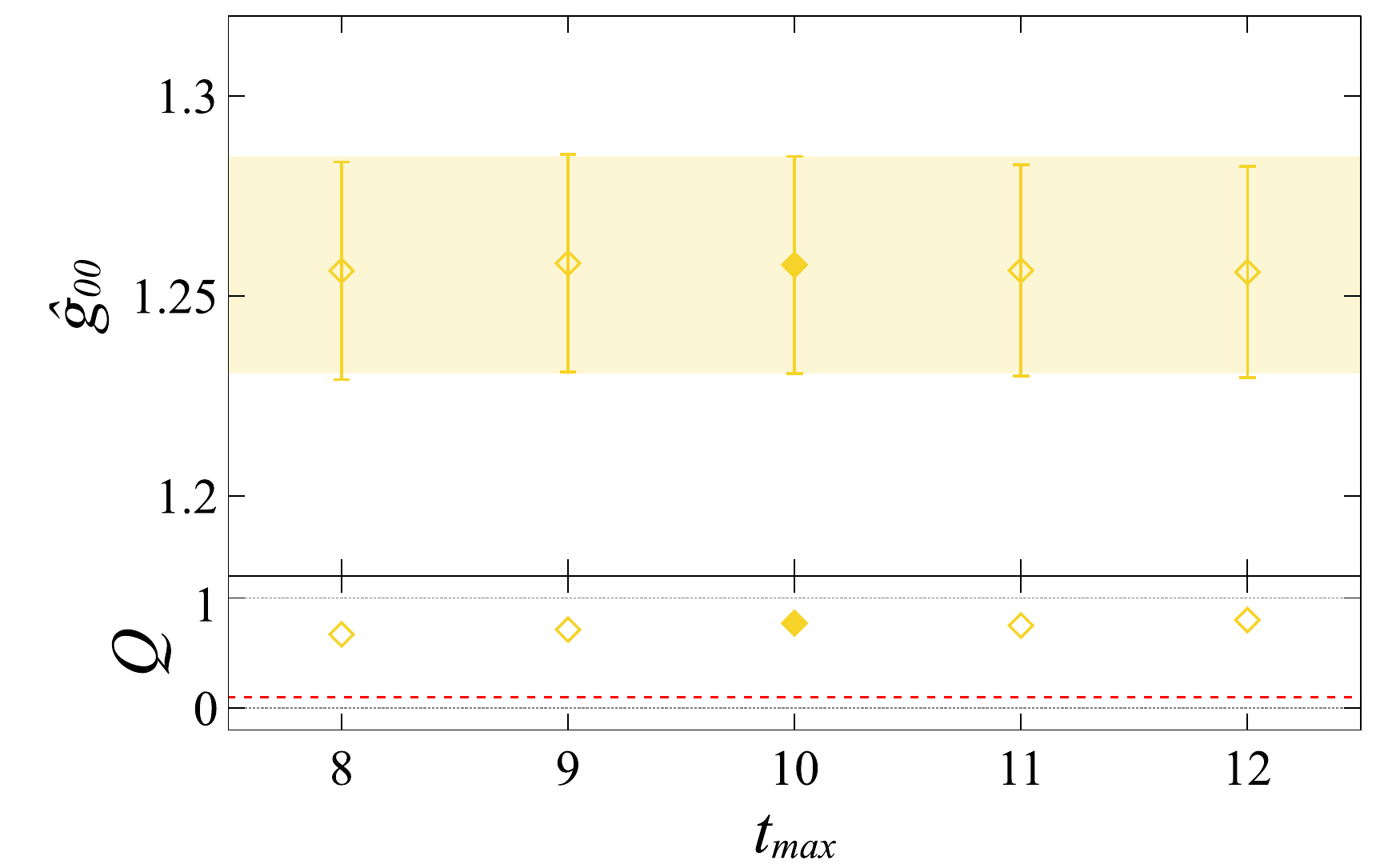}
	\caption{Stability plot under varying $t_{max}$ for the seven-state fit with $t_{min}=3$. The solid symbol denotes the preferred two-point correlator fit.}
	\label{fig:3pt_stability_tmax}
\end{figure}

The stability of $\hat{g}_{00}$ under varying the fit region and number of excited states is shown in Figs.~\ref{fig:3pt_stability} and \ref{fig:3pt_stability_tmax}. 
For all simultaneous two- and three-point fits, we perform the preferred two-point fit as discussed in Sec.~\ref{sec:2ptcorr}.

For fits with $t_{min}=2, 3$, Fig.~\ref{fig:3pt_stability} shows stability under varying the number of states for fits with more than five states, while the Bayes factor prefers the five-state fit (solid yellow diamonds). For fits with $t_{min} \geq 4$, we also observe stability; however the Bayes factor prefers fits with less parameters ($<$ four states) over fits that qualitatively look more stable ($\geq$ five states), suggesting that there is not enough data to support fits with a large number of states. The fit at $t_{min}=1$ is stable after seven states; however Eq.~(\ref{eq:contactcoeff}) shows that $t=1$ includes no information about $g_{00}$, the parameter of interest. Therefore for this study, we choose the five-state fit with $t=[2, 10]$ as the preferred fit. The preferred fit is shown to be insensitive under variations of $t_{max}$ as demonstrated in Fig.~\ref{fig:3pt_stability_tmax}.

The posterior distributions from the simultaneous fit are used to reconstruct the fitted curve and are shown by the green bands in Figs.~\ref{fig:dmeff} and \ref{fig:dmeff_ps}. Similar to the effective mass and scaled correlator, we recover $g_{00}$ in the $t\rightarrow\infty$ limit, as shown by the exact overlap of $\hat{g}_{00}$ and the green fit curve in the large-$t$ limit. 
Following this analysis we quote the value of the bare axial charge
\begin{equation}
\mathring{g}_{A} = 1.258(27)
\end{equation}
on the $a\approx 0.15$~fm lattice with $m_\pi \approx 310$~MeV.
Determining the axial renormalization factor requires calculations at multiple quark masses to allow for an extrapolation to the chiral limit, which we do not perform in this work.
These calculations use MDWF fermions, allowing for a determination of $Z_A$ by computing the pion decay constant with both the conserved five-dimensional axial Ward identity and with the nonconserved four-dimensional current.  We find for this ensemble
\begin{equation}
Z_A = 0.9646(6)\, .
\end{equation}
While this is a quark-mass-dependent renormalization, we can still compare our renormalized value of $g_A$ with that determined using clover valence quarks on similar HISQ ensembles in Ref.~\cite{Bhattacharya:2016zcn}. They found $g_A = 1.221(28)$ at $m_\pi\approx 310$~MeV with $a\approx0.12$~fm, as compared to our value of
\begin{equation}
	Z_A\, \mathring{g}_A 
	= 1.213(26)\, .
\end{equation}
A more comprehensive and direct comparison with the results of Ref.~\cite{Bhattacharya:2016zcn} will be made in a forthcoming publication~\cite{Berkowitz:2017gql}.
When comparing on specific ensembles, it is observed that roughly an order of magnitude less inversions are required to achieve the same statistical precision with this new method as compared to the fixed source-sink separation method. 

\section{Generalizations\label{Sec:4}}

The generalization of this method to nonzero momentum transfer and/or flavor-changing currents is straightforward. To inject momentum with the current, instead of summing over a constant operator, the Feynman-Hellmann propagator would be constructed with a momentum phase
\begin{equation}
S^\G_q(y,x) = \sum_{z=(t_z,\mathbf{z})} e^{iq\cdot z}S(y,z) \G S(z,x)\, .
\end{equation} 
With the current implementation, each choice of momentum injection would require the computation of an additional Feynman-Hellmann propagator.

More complicated space-time-dependent current insertions can also be considered. The modification of the numerator function is also straightforward.
Unlike Eq.~\eqref{eq:N1_intermediate}, the energy of the incoming and outgoing states would no longer be equal, $E_m(p) \neq E_n(p+q)$ for any of the states $|n\rangle$ and $|m\rangle$. The exception would be if the calculation is performed in the Breit frame when $\mathbf{p+q} = -\mathbf{p}$. In this case, the numerator expression from region I would still be given by Eq.~\eqref{eq:NI}.  Otherwise, the entire contribution from region I is parametrized by the second term in Eq.~\eqref{eq:NI} where the state labels $n$ and $m$ now carry information about the momentum as well.

Flavor-changing interactions are just as straightforward. In this case, the Feynman-Hellmann propagator is constructed as
\begin{equation}
S^\G_{j\leftarrow i}(y,x) = \sum_{z=(t_z,\mathbf{z})} S_j(y,z) \G S_i(z,x)\, ,
\end{equation} 
for a flavor-changing interaction $i \rightarrow j$, with a trivial generalization to include momentum transfer as well. As with nonzero momentum transfer, the numerator expression from region I simplifies to just the second line of Eq.~\eqref{eq:NI} as the energies of the incoming and outgoing states will not match.

This method can also be extended to consider two current insertions. 
A feasible technique for calculating two-current insertion matrix elements is needed to permit the LQCD evaluation of, e.g., two-photon corrections to the Lamb shift in muonic helium-3 ions~\cite{Carlson:2016cii}, which may shed light on the proton radius problem, and the $\gamma$-$Z$ box diagram corrections to electromagnetic structure functions~\cite{Rislow:2013vta} needed to improve the interpretation of results from the $Q_{\it weak}$ experiment~\cite{PhysRevLett.111.141803}.

We first generalize Eqs.~(\ref{eq:twopt_l})--(\ref{eq:Z_lam}) to include multiple currents $\mathcal{J}_i$ with associated couplings $\l_i$. In the presence of multiple currents, the modified action of Eq.~(\ref{eq:action_l}) is
\begin{equation}
S_{\mathbf{\l}} = \bm{\l} \cdot \int dt\, \boldsymbol{\mathcal{J}}(t)\, .
\end{equation}
Two-current-insertion matrix elements can then be associated with the second derivative of the effective mass,
\begin{align}\label{eq:d2meff}
	\frac{\partial^2 m^{\it eff}_{\mathbf{\l}}(t,\t)}{\partial \l_i \partial \l_j} \bigg|_{\mathbf{\l}=\mathbf{0}}
	&= \frac{1}{\t} \bigg[ R_{ij}(t+\t) - R_{ij}(t)  \\
	&- R_i(t+\t)R_j(t+\t) + R_i(t) R_j(t) \bigg].\nonumber
\end{align}
where $R_i$ is a generalization of Eq.~(\ref{eq:R(t)}),
\begin{equation}
R_i(t) \equiv \frac{\int dt^\prime \langle \O | T\{ \mc{O}(t) \mathcal{J}_i(t^\prime) \mc{O}^\dagger(0) \} | \O \rangle}{C(t)}\, ,
\end{equation} 
and $R_{ij}$ is the ratio of the two-current insertion matrix element to the two-point function,
\begin{equation}
R_{ij}(t) \equiv \frac{\int dt^\prime dt^{\prime\prime} \langle \O | T\{ \mc{O}(t) \mathcal{J}_i(t^\prime) \mathcal{J}_j(t^{\prime\prime}) \mc{O}^\dagger(0) \} | \O \rangle}{C(t)} .
\end{equation} 
Like the first derivative of the effective mass in Eq.~(\ref{eq:dmeff}), the second derivative of the effective mass (\textit{i.e.} Eq.~(\ref{eq:d2meff})) benefits from an exact cancellation of the vacuum matrix elements of $\mathcal{J}_i$, $\mathcal{J}_j$, and $\mathcal{J}_i \mathcal{J}_j$. The spectral decomposition of Eq.~(\ref{eq:d2meff}) precedes in analogy with that outlined in Sec.~\ref{sec:new_method}, albeit with a few added complications that we intend to address in future work.

\section{Conclusions and Outlook\label{Sec:5}}

This work presents a computationally efficient and comprehensive implementation of lattice calculations of hadronic matrix elements utilizing ideas which can be related to the Feynman-Hellmann Theorem. In particular, the derivative correlation function was analytically derived such that a background field is not explicitly needed, reducing the overall number of propagator solves required to determine the linear response of the theory to an external current.

The example calculation of the nucleon axial charge demonstrates that from the calculation of a low-statistics run, we are able to achieve a 2.1\% uncertainty on $g_A$, including all systematic errors from the fitting procedure. The derivation of systematic effects associated with this method, including excited-state contamination, mesonic propagating modes, and signal from contact operators were completely presented up to terms suppressed by $\mc{O}(e^{-E_nT})$, where corrections to include higher-order thermal effect are self-evident. Access to all source-sink separations also allows us to leverage knowledge of the complete spectral decomposition and demonstrate complete control over all systematic uncertainties originating from the fitting procedure. The benefits of increased stochastic sampling is made possible by 1) the sum over the current insertion time $t^\prime$ leading to $\mathrm{O}(t)$ increase in statistics and 2) the ability to explore fits with multiple source-sink separations, allowing for the eventual \textit{exponential} gain in the signal-over-noise ratio due to fitting to a source-sink separation of $\approx 0.45$~fm ($t_{min}=3$).  Consequently, we find approximately an order of magnitude less propagator solves are required to achieve the same stochastic precision when calculating a single matrix elements compared to calculations that use the standard fixed source-sink separation method. However, a new propagator solve is required for each matrix element of interest, including for each momentum insertion of the current. 

The generalization of this method to nonzero momentum transfer, flavor-changing currents, and multiple current insertions is straightforward. The ideas presented here are not restricted to lattice field theory calculations, but applicable to any theory with the notion of a bilinear current.

In summary, the method we presented sacrifices the flexibility of calculating the matrix element of any operator at any momentum in exchange for the determination of the matrix element of one operator at fixed momentum with precision that is one computing generation ahead of its time.

\acknowledgments

We thank T. Izubuchi for discussions. The LQCD calculations for this work utilized a modification of the \texttt{Chroma} software suite~\cite{Edwards:2004sx}, the ``irresponsibly fast'' QUDA library~\cite{Clark:2009wm,Babich:2011np} and the the highly efficient HDF5 I/O Library~\cite{hdf5} with an interface to HDF5 in the USQCD QDP++ package that was added with CalLat SciDAC 3 support~\cite{Kurth:2015mqa}. The calculations were performed at the Jefferson Lab High Performance Computing Center on facilities of the USQCD Collaboration, which are funded by the Office of Science of the U.S. Department of Energy, 
and at LLNL through a Multiprogrammatic and Institutional Computing program for the Grand Challenge allocation.

This research was supported in part by The U.S. Department of Energy, Office of Science; under contract number DE-AC05-06OR23177, which Jefferson Science Associates, LLC, manages and operates the Jefferson Lab (KNO, AWL);
under contract number DE-AC02-05CH11231, which the Regents of the University of California manage and operate Lawrence Berkeley National Laboratory and the National Energy Research Scientific Computing Center (CCC, TK, AWL);
Office of Nuclear Physics under Grant Number DE-FG02-04ER41302 (CMB, KNO), 
and the Double-Beta Decay Topical Collaboration under Contract number DE-SC0015376 (AWL); Office of Advanced Scientific Computing Research, Scientific Discovery through Advanced Computing (SciDAC) program under Award Number KB0301052 (TK, AWL); The Office of Nuclear Physics The DOE Early Career Research Program, Office of Nuclear Physics under FWP Number NQCDAWL (CCC, AWL).

\bibliography{./c51_bib}

\begin{thebibliography}{58}%
\makeatletter
\providecommand \@ifxundefined [1]{%
 \@ifx{#1\undefined}
}%
\providecommand \@ifnum [1]{%
 \ifnum #1\expandafter \@firstoftwo
 \else \expandafter \@secondoftwo
 \fi
}%
\providecommand \@ifx [1]{%
 \ifx #1\expandafter \@firstoftwo
 \else \expandafter \@secondoftwo
 \fi
}%
\providecommand \natexlab [1]{#1}%
\providecommand \enquote  [1]{``#1''}%
\providecommand \bibnamefont  [1]{#1}%
\providecommand \bibfnamefont [1]{#1}%
\providecommand \citenamefont [1]{#1}%
\providecommand \href@noop [0]{\@secondoftwo}%
\providecommand \href [0]{\begingroup \@sanitize@url \@href}%
\providecommand \@href[1]{\@@startlink{#1}\@@href}%
\providecommand \@@href[1]{\endgroup#1\@@endlink}%
\providecommand \@sanitize@url [0]{\catcode `\\12\catcode `\$12\catcode
  `\&12\catcode `\#12\catcode `\^12\catcode `\_12\catcode `\%12\relax}%
\providecommand \@@startlink[1]{}%
\providecommand \@@endlink[0]{}%
\providecommand \url  [0]{\begingroup\@sanitize@url \@url }%
\providecommand \@url [1]{\endgroup\@href {#1}{\urlprefix }}%
\providecommand \urlprefix  [0]{URL }%
\providecommand \Eprint [0]{\href }%
\providecommand \doibase [0]{http://dx.doi.org/}%
\providecommand \selectlanguage [0]{\@gobble}%
\providecommand \bibinfo  [0]{\@secondoftwo}%
\providecommand \bibfield  [0]{\@secondoftwo}%
\providecommand \translation [1]{[#1]}%
\providecommand \BibitemOpen [0]{}%
\providecommand \bibitemStop [0]{}%
\providecommand \bibitemNoStop [0]{.\EOS\space}%
\providecommand \EOS [0]{\spacefactor3000\relax}%
\providecommand \BibitemShut  [1]{\csname bibitem#1\endcsname}%
\let\auto@bib@innerbib\@empty
\bibitem [{\citenamefont {{G{\"u}ttinger, P.}}(1932)}]{Guettinger1932}%
  \BibitemOpen
  \bibfield  {author} {\bibinfo {author} {\bibnamefont {{G{\"u}ttinger, P.}}},\
  }\bibfield  {title} {\enquote {\bibinfo {title} {Das verhalten von atomen im
  magnetischen drehfeld},}\ }\href {\doibase 10.1007/BF01351211} {\bibfield
  {journal} {\bibinfo  {journal} {{Zeitschrift f{\"u}r Physik}}\ }\textbf
  {\bibinfo {volume} {73}},\ \bibinfo {pages} {169--184} (\bibinfo {year}
  {1932})}\BibitemShut {NoStop}%
\bibitem [{\citenamefont {Pauli}(1933)}]{Pauli1933}%
  \BibitemOpen
  \bibfield  {author} {\bibinfo {author} {\bibfnamefont {W.}~\bibnamefont
  {Pauli}},\ }\href@noop {} {\emph {\bibinfo {title} {Principles of Wave
  Mechanics}}},\ Vol.~\bibinfo {volume} {24}\ (\bibinfo  {publisher}
  {Springer},\ \bibinfo {year} {1933})\ p.\ \bibinfo {pages} {162}\BibitemShut
  {NoStop}%
\bibitem [{\citenamefont {Hellmann}(1937)}]{Hellmann1933}%
  \BibitemOpen
  \bibfield  {author} {\bibinfo {author} {\bibfnamefont {H}~\bibnamefont
  {Hellmann}},\ }\href@noop {} {\emph {\bibinfo {title} {Einf\"{u}hrung in die
  Quantenchemie}}}\ (\bibinfo  {publisher} {Franz Deuticke},\ \bibinfo {year}
  {1937})\ p.\ \bibinfo {pages} {285},\ \bibinfo {note}
  {{OL21481721M}}\BibitemShut {NoStop}%
\bibitem [{\citenamefont {Feynman}(1939)}]{PhysRev.56.340}%
  \BibitemOpen
  \bibfield  {author} {\bibinfo {author} {\bibfnamefont {R.~P.}\ \bibnamefont
  {Feynman}},\ }\bibfield  {title} {\enquote {\bibinfo {title} {Forces in
  molecules},}\ }\href {\doibase 10.1103/PhysRev.56.340} {\bibfield  {journal}
  {\bibinfo  {journal} {Phys. Rev.}\ }\textbf {\bibinfo {volume} {56}},\
  \bibinfo {pages} {340--343} (\bibinfo {year} {1939})}\BibitemShut {NoStop}%
\bibitem [{\citenamefont {Procura}\ \emph {et~al.}(2004)\citenamefont
  {Procura}, \citenamefont {Hemmert},\ and\ \citenamefont
  {Weise}}]{Procura:2003ig}%
  \BibitemOpen
  \bibfield  {author} {\bibinfo {author} {\bibfnamefont {Massimiliano}\
  \bibnamefont {Procura}}, \bibinfo {author} {\bibfnamefont {Thomas~R.}\
  \bibnamefont {Hemmert}}, \ and\ \bibinfo {author} {\bibfnamefont {Wolfram}\
  \bibnamefont {Weise}},\ }\bibfield  {title} {\enquote {\bibinfo {title}
  {{Nucleon mass, sigma term and lattice QCD}},}\ }\href {\doibase
  10.1103/PhysRevD.69.034505} {\bibfield  {journal} {\bibinfo  {journal} {Phys.
  Rev.}\ }\textbf {\bibinfo {volume} {D69}},\ \bibinfo {pages} {034505}
  (\bibinfo {year} {2004})},\ \Eprint {http://arxiv.org/abs/hep-lat/0309020}
  {arXiv:hep-lat/0309020 [hep-lat]} \BibitemShut {NoStop}%
\bibitem [{\citenamefont {Procura}\ \emph {et~al.}(2006)\citenamefont
  {Procura}, \citenamefont {Musch}, \citenamefont {Wollenweber}, \citenamefont
  {Hemmert},\ and\ \citenamefont {Weise}}]{Procura:2006bj}%
  \BibitemOpen
  \bibfield  {author} {\bibinfo {author} {\bibfnamefont {M.}~\bibnamefont
  {Procura}}, \bibinfo {author} {\bibfnamefont {B.~U.}\ \bibnamefont {Musch}},
  \bibinfo {author} {\bibfnamefont {T.}~\bibnamefont {Wollenweber}}, \bibinfo
  {author} {\bibfnamefont {T.~R.}\ \bibnamefont {Hemmert}}, \ and\ \bibinfo
  {author} {\bibfnamefont {W.}~\bibnamefont {Weise}},\ }\bibfield  {title}
  {\enquote {\bibinfo {title} {{Nucleon mass: From lattice QCD to the chiral
  limit}},}\ }\href {\doibase 10.1103/PhysRevD.73.114510} {\bibfield  {journal}
  {\bibinfo  {journal} {Phys. Rev.}\ }\textbf {\bibinfo {volume} {D73}},\
  \bibinfo {pages} {114510} (\bibinfo {year} {2006})},\ \Eprint
  {http://arxiv.org/abs/hep-lat/0603001} {arXiv:hep-lat/0603001 [hep-lat]}
  \BibitemShut {NoStop}%
\bibitem [{\citenamefont {Alexandrou}\ \emph {et~al.}(2008)\citenamefont
  {Alexandrou} \emph {et~al.}}]{Alexandrou:2008tn}%
  \BibitemOpen
  \bibfield  {author} {\bibinfo {author} {\bibfnamefont {C.}~\bibnamefont
  {Alexandrou}} \emph {et~al.} (\bibinfo {collaboration} {European Twisted
  Mass}),\ }\bibfield  {title} {\enquote {\bibinfo {title} {{Light baryon
  masses with dynamical twisted mass fermions}},}\ }\href {\doibase
  10.1103/PhysRevD.78.014509} {\bibfield  {journal} {\bibinfo  {journal} {Phys.
  Rev.}\ }\textbf {\bibinfo {volume} {D78}},\ \bibinfo {pages} {014509}
  (\bibinfo {year} {2008})},\ \Eprint {http://arxiv.org/abs/0803.3190}
  {arXiv:0803.3190 [hep-lat]} \BibitemShut {NoStop}%
\bibitem [{\citenamefont {Walker-Loud}\ \emph {et~al.}(2009)\citenamefont
  {Walker-Loud}, \citenamefont {Lin}, \citenamefont {Richards}, \citenamefont
  {Edwards}, \citenamefont {Engelhardt} \emph {et~al.}}]{WalkerLoud:2008bp}%
  \BibitemOpen
  \bibfield  {author} {\bibinfo {author} {\bibfnamefont {A.}~\bibnamefont
  {Walker-Loud}}, \bibinfo {author} {\bibfnamefont {H.-W.}\ \bibnamefont
  {Lin}}, \bibinfo {author} {\bibfnamefont {D.G.}\ \bibnamefont {Richards}},
  \bibinfo {author} {\bibfnamefont {R.G.}\ \bibnamefont {Edwards}}, \bibinfo
  {author} {\bibfnamefont {M.}~\bibnamefont {Engelhardt}},  \emph {et~al.},\
  }\bibfield  {title} {\enquote {\bibinfo {title} {{Light hadron spectroscopy
  using domain wall valence quarks on an Asqtad sea}},}\ }\href {\doibase
  10.1103/PhysRevD.79.054502} {\bibfield  {journal} {\bibinfo  {journal}
  {Phys.Rev.}\ }\textbf {\bibinfo {volume} {D79}},\ \bibinfo {pages} {054502}
  (\bibinfo {year} {2009})},\ \Eprint {http://arxiv.org/abs/0806.4549}
  {arXiv:0806.4549 [hep-lat]} \BibitemShut {NoStop}%
\bibitem [{\citenamefont {Ohki}\ \emph {et~al.}(2008)\citenamefont {Ohki},
  \citenamefont {Fukaya}, \citenamefont {Hashimoto}, \citenamefont {Kaneko},
  \citenamefont {Matsufuru}, \citenamefont {Noaki}, \citenamefont {Onogi},
  \citenamefont {Shintani},\ and\ \citenamefont {Yamada}}]{Ohki:2008ff}%
  \BibitemOpen
  \bibfield  {author} {\bibinfo {author} {\bibfnamefont {H.}~\bibnamefont
  {Ohki}}, \bibinfo {author} {\bibfnamefont {H.}~\bibnamefont {Fukaya}},
  \bibinfo {author} {\bibfnamefont {S.}~\bibnamefont {Hashimoto}}, \bibinfo
  {author} {\bibfnamefont {T.}~\bibnamefont {Kaneko}}, \bibinfo {author}
  {\bibfnamefont {H.}~\bibnamefont {Matsufuru}}, \bibinfo {author}
  {\bibfnamefont {J.}~\bibnamefont {Noaki}}, \bibinfo {author} {\bibfnamefont
  {T.}~\bibnamefont {Onogi}}, \bibinfo {author} {\bibfnamefont
  {E.}~\bibnamefont {Shintani}}, \ and\ \bibinfo {author} {\bibfnamefont
  {N.}~\bibnamefont {Yamada}},\ }\bibfield  {title} {\enquote {\bibinfo {title}
  {{Nucleon sigma term and strange quark content from lattice QCD with exact
  chiral symmetry}},}\ }\href {\doibase 10.1103/PhysRevD.78.054502} {\bibfield
  {journal} {\bibinfo  {journal} {Phys. Rev.}\ }\textbf {\bibinfo {volume}
  {D78}},\ \bibinfo {pages} {054502} (\bibinfo {year} {2008})},\ \Eprint
  {http://arxiv.org/abs/0806.4744} {arXiv:0806.4744 [hep-lat]} \BibitemShut
  {NoStop}%
\bibitem [{\citenamefont {Young}\ and\ \citenamefont
  {Thomas}(2010)}]{Young:2009zb}%
  \BibitemOpen
  \bibfield  {author} {\bibinfo {author} {\bibfnamefont {R.~D.}\ \bibnamefont
  {Young}}\ and\ \bibinfo {author} {\bibfnamefont {A.~W.}\ \bibnamefont
  {Thomas}},\ }\bibfield  {title} {\enquote {\bibinfo {title} {{Octet baryon
  masses and sigma terms from an SU(3) chiral extrapolation}},}\ }\href
  {\doibase 10.1103/PhysRevD.81.014503} {\bibfield  {journal} {\bibinfo
  {journal} {Phys. Rev.}\ }\textbf {\bibinfo {volume} {D81}},\ \bibinfo {pages}
  {014503} (\bibinfo {year} {2010})},\ \Eprint {http://arxiv.org/abs/0901.3310}
  {arXiv:0901.3310 [hep-lat]} \BibitemShut {NoStop}%
\bibitem [{\citenamefont {Durr}\ \emph {et~al.}(2012)\citenamefont {Durr} \emph
  {et~al.}}]{Durr:2011mp}%
  \BibitemOpen
  \bibfield  {author} {\bibinfo {author} {\bibfnamefont {S.}~\bibnamefont
  {Durr}} \emph {et~al.},\ }\bibfield  {title} {\enquote {\bibinfo {title}
  {{Sigma term and strangeness content of octet baryons}},}\ }\href {\doibase
  10.1103/PhysRevD.85.014509, 10.1103/PhysRevD.93.039905} {\bibfield  {journal}
  {\bibinfo  {journal} {Phys. Rev.}\ }\textbf {\bibinfo {volume} {D85}},\
  \bibinfo {pages} {014509} (\bibinfo {year} {2012})},\ \bibinfo {note}
  {[Erratum: Phys. Rev.D93,no.3,039905(2016)]},\ \Eprint
  {http://arxiv.org/abs/1109.4265} {arXiv:1109.4265 [hep-lat]} \BibitemShut
  {NoStop}%
\bibitem [{\citenamefont {Horsley}\ \emph {et~al.}(2012)\citenamefont
  {Horsley}, \citenamefont {Nakamura}, \citenamefont {Perlt}, \citenamefont
  {Pleiter}, \citenamefont {Rakow}, \citenamefont {Schierholz}, \citenamefont
  {Schiller}, \citenamefont {Stuben}, \citenamefont {Winter},\ and\
  \citenamefont {Zanotti}}]{Horsley:2011wr}%
  \BibitemOpen
  \bibfield  {author} {\bibinfo {author} {\bibfnamefont {R.}~\bibnamefont
  {Horsley}}, \bibinfo {author} {\bibfnamefont {Y.}~\bibnamefont {Nakamura}},
  \bibinfo {author} {\bibfnamefont {H.}~\bibnamefont {Perlt}}, \bibinfo
  {author} {\bibfnamefont {D.}~\bibnamefont {Pleiter}}, \bibinfo {author}
  {\bibfnamefont {P.~E.~L.}\ \bibnamefont {Rakow}}, \bibinfo {author}
  {\bibfnamefont {G.}~\bibnamefont {Schierholz}}, \bibinfo {author}
  {\bibfnamefont {A.}~\bibnamefont {Schiller}}, \bibinfo {author}
  {\bibfnamefont {H.}~\bibnamefont {Stuben}}, \bibinfo {author} {\bibfnamefont
  {F.}~\bibnamefont {Winter}}, \ and\ \bibinfo {author} {\bibfnamefont {J.~M.}\
  \bibnamefont {Zanotti}} (\bibinfo {collaboration} {QCDSF-UKQCD}),\ }\bibfield
   {title} {\enquote {\bibinfo {title} {{Hyperon sigma terms for 2+1 quark
  flavours}},}\ }\href {\doibase 10.1103/PhysRevD.85.034506} {\bibfield
  {journal} {\bibinfo  {journal} {Phys. Rev.}\ }\textbf {\bibinfo {volume}
  {D85}},\ \bibinfo {pages} {034506} (\bibinfo {year} {2012})},\ \Eprint
  {http://arxiv.org/abs/1110.4971} {arXiv:1110.4971 [hep-lat]} \BibitemShut
  {NoStop}%
\bibitem [{\citenamefont {Freeman}\ and\ \citenamefont
  {Toussaint}(2013)}]{Freeman:2012ry}%
  \BibitemOpen
  \bibfield  {author} {\bibinfo {author} {\bibfnamefont {Walter}\ \bibnamefont
  {Freeman}}\ and\ \bibinfo {author} {\bibfnamefont {Doug}\ \bibnamefont
  {Toussaint}} (\bibinfo {collaboration} {MILC}),\ }\bibfield  {title}
  {\enquote {\bibinfo {title} {{Intrinsic strangeness and charm of the nucleon
  using improved staggered fermions}},}\ }\href {\doibase
  10.1103/PhysRevD.88.054503} {\bibfield  {journal} {\bibinfo  {journal} {Phys.
  Rev.}\ }\textbf {\bibinfo {volume} {D88}},\ \bibinfo {pages} {054503}
  (\bibinfo {year} {2013})},\ \Eprint {http://arxiv.org/abs/1204.3866}
  {arXiv:1204.3866 [hep-lat]} \BibitemShut {NoStop}%
\bibitem [{\citenamefont {Bali}\ \emph {et~al.}(2013)\citenamefont {Bali} \emph
  {et~al.}}]{Bali:2012qs}%
  \BibitemOpen
  \bibfield  {author} {\bibinfo {author} {\bibfnamefont {G.~S.}\ \bibnamefont
  {Bali}} \emph {et~al.},\ }\bibfield  {title} {\enquote {\bibinfo {title}
  {{Nucleon mass and sigma term from lattice QCD with two light fermion
  flavors}},}\ }\href {\doibase 10.1016/j.nuclphysb.2012.08.009} {\bibfield
  {journal} {\bibinfo  {journal} {Nucl. Phys.}\ }\textbf {\bibinfo {volume}
  {B866}},\ \bibinfo {pages} {1--25} (\bibinfo {year} {2013})},\ \Eprint
  {http://arxiv.org/abs/1206.7034} {arXiv:1206.7034 [hep-lat]} \BibitemShut
  {NoStop}%
\bibitem [{\citenamefont {Ohki}\ \emph {et~al.}(2013)\citenamefont {Ohki},
  \citenamefont {Takeda}, \citenamefont {Aoki}, \citenamefont {Hashimoto},
  \citenamefont {Kaneko}, \citenamefont {Matsufuru}, \citenamefont {Noaki},\
  and\ \citenamefont {Onogi}}]{Oksuzian:2012rzb}%
  \BibitemOpen
  \bibfield  {author} {\bibinfo {author} {\bibfnamefont {H.}~\bibnamefont
  {Ohki}}, \bibinfo {author} {\bibfnamefont {K.}~\bibnamefont {Takeda}},
  \bibinfo {author} {\bibfnamefont {S.}~\bibnamefont {Aoki}}, \bibinfo {author}
  {\bibfnamefont {S.}~\bibnamefont {Hashimoto}}, \bibinfo {author}
  {\bibfnamefont {T.}~\bibnamefont {Kaneko}}, \bibinfo {author} {\bibfnamefont
  {H.}~\bibnamefont {Matsufuru}}, \bibinfo {author} {\bibfnamefont
  {J.}~\bibnamefont {Noaki}}, \ and\ \bibinfo {author} {\bibfnamefont
  {T.}~\bibnamefont {Onogi}} (\bibinfo {collaboration} {JLQCD}),\ }\bibfield
  {title} {\enquote {\bibinfo {title} {{Nucleon strange quark content from
  $N_f=2+1$ lattice QCD with exact chiral symmetry}},}\ }\href {\doibase
  10.1103/PhysRevD.87.034509} {\bibfield  {journal} {\bibinfo  {journal} {Phys.
  Rev.}\ }\textbf {\bibinfo {volume} {D87}},\ \bibinfo {pages} {034509}
  (\bibinfo {year} {2013})},\ \Eprint {http://arxiv.org/abs/1208.4185}
  {arXiv:1208.4185 [hep-lat]} \BibitemShut {NoStop}%
\bibitem [{\citenamefont {Semke}\ and\ \citenamefont
  {Lutz}(2012)}]{Semke:2012gs}%
  \BibitemOpen
  \bibfield  {author} {\bibinfo {author} {\bibfnamefont {A.}~\bibnamefont
  {Semke}}\ and\ \bibinfo {author} {\bibfnamefont {M.~F.~M.}\ \bibnamefont
  {Lutz}},\ }\bibfield  {title} {\enquote {\bibinfo {title} {{Strangeness in
  the baryon ground states}},}\ }\href {\doibase
  10.1016/j.physletb.2012.09.008} {\bibfield  {journal} {\bibinfo  {journal}
  {Phys. Lett.}\ }\textbf {\bibinfo {volume} {B717}},\ \bibinfo {pages}
  {242--247} (\bibinfo {year} {2012})},\ \Eprint
  {http://arxiv.org/abs/1202.3556} {arXiv:1202.3556 [hep-ph]} \BibitemShut
  {NoStop}%
\bibitem [{\citenamefont {Shanahan}\ \emph {et~al.}(2013)\citenamefont
  {Shanahan}, \citenamefont {Thomas},\ and\ \citenamefont
  {Young}}]{Shanahan:2012wh}%
  \BibitemOpen
  \bibfield  {author} {\bibinfo {author} {\bibfnamefont {P.~E.}\ \bibnamefont
  {Shanahan}}, \bibinfo {author} {\bibfnamefont {A.~W.}\ \bibnamefont
  {Thomas}}, \ and\ \bibinfo {author} {\bibfnamefont {R.~D.}\ \bibnamefont
  {Young}},\ }\bibfield  {title} {\enquote {\bibinfo {title} {{Sigma terms from
  an SU(3) chiral extrapolation}},}\ }\href {\doibase
  10.1103/PhysRevD.87.074503} {\bibfield  {journal} {\bibinfo  {journal} {Phys.
  Rev.}\ }\textbf {\bibinfo {volume} {D87}},\ \bibinfo {pages} {074503}
  (\bibinfo {year} {2013})},\ \Eprint {http://arxiv.org/abs/1205.5365}
  {arXiv:1205.5365 [nucl-th]} \BibitemShut {NoStop}%
\bibitem [{\citenamefont {Ren}\ \emph {et~al.}(2012)\citenamefont {Ren},
  \citenamefont {Geng}, \citenamefont {Martin~Camalich}, \citenamefont {Meng},\
  and\ \citenamefont {Toki}}]{Ren:2012aj}%
  \BibitemOpen
  \bibfield  {author} {\bibinfo {author} {\bibfnamefont {X.~L.}\ \bibnamefont
  {Ren}}, \bibinfo {author} {\bibfnamefont {L.~S.}\ \bibnamefont {Geng}},
  \bibinfo {author} {\bibfnamefont {J.}~\bibnamefont {Martin~Camalich}},
  \bibinfo {author} {\bibfnamefont {J.}~\bibnamefont {Meng}}, \ and\ \bibinfo
  {author} {\bibfnamefont {H.}~\bibnamefont {Toki}},\ }\bibfield  {title}
  {\enquote {\bibinfo {title} {{Octet baryon masses in
  next-to-next-to-next-to-leading order covariant baryon chiral perturbation
  theory}},}\ }\href {\doibase 10.1007/JHEP12(2012)073} {\bibfield  {journal}
  {\bibinfo  {journal} {JHEP}\ }\textbf {\bibinfo {volume} {12}},\ \bibinfo
  {pages} {073} (\bibinfo {year} {2012})},\ \Eprint
  {http://arxiv.org/abs/1209.3641} {arXiv:1209.3641 [nucl-th]} \BibitemShut
  {NoStop}%
\bibitem [{\citenamefont {Junnarkar}\ and\ \citenamefont
  {Walker-Loud}(2013)}]{Junnarkar:2013ac}%
  \BibitemOpen
  \bibfield  {author} {\bibinfo {author} {\bibfnamefont {Parikshit}\
  \bibnamefont {Junnarkar}}\ and\ \bibinfo {author} {\bibfnamefont {Andre}\
  \bibnamefont {Walker-Loud}},\ }\bibfield  {title} {\enquote {\bibinfo {title}
  {{The Scalar Strange Content of the Nucleon from Lattice QCD}},}\ }\href
  {\doibase 10.1103/PhysRevD.87.114510} {\bibfield  {journal} {\bibinfo
  {journal} {Phys.Rev.}\ }\textbf {\bibinfo {volume} {D87}},\ \bibinfo {pages}
  {114510} (\bibinfo {year} {2013})},\ \Eprint {http://arxiv.org/abs/1301.1114}
  {arXiv:1301.1114 [hep-lat]} \BibitemShut {NoStop}%
\bibitem [{\citenamefont {Durr}\ \emph {et~al.}(2016)\citenamefont {Durr} \emph
  {et~al.}}]{Durr:2015dna}%
  \BibitemOpen
  \bibfield  {author} {\bibinfo {author} {\bibfnamefont {S.}~\bibnamefont
  {Durr}} \emph {et~al.},\ }\bibfield  {title} {\enquote {\bibinfo {title}
  {{Lattice computation of the nucleon scalar quark contents at the physical
  point}},}\ }\href {\doibase 10.1103/PhysRevLett.116.172001} {\bibfield
  {journal} {\bibinfo  {journal} {Phys. Rev. Lett.}\ }\textbf {\bibinfo
  {volume} {116}},\ \bibinfo {pages} {172001} (\bibinfo {year} {2016})},\
  \Eprint {http://arxiv.org/abs/1510.08013} {arXiv:1510.08013 [hep-lat]}
  \BibitemShut {NoStop}%
\bibitem [{\citenamefont {Hill}\ and\ \citenamefont
  {Solon}(2014)}]{Hill:2013hoa}%
  \BibitemOpen
  \bibfield  {author} {\bibinfo {author} {\bibfnamefont {Richard~J.}\
  \bibnamefont {Hill}}\ and\ \bibinfo {author} {\bibfnamefont {Mikhail~P.}\
  \bibnamefont {Solon}},\ }\bibfield  {title} {\enquote {\bibinfo {title}
  {{WIMP-nucleon scattering with heavy WIMP effective theory}},}\ }\href
  {\doibase 10.1103/PhysRevLett.112.211602} {\bibfield  {journal} {\bibinfo
  {journal} {Phys. Rev. Lett.}\ }\textbf {\bibinfo {volume} {112}},\ \bibinfo
  {pages} {211602} (\bibinfo {year} {2014})},\ \Eprint
  {http://arxiv.org/abs/1309.4092} {arXiv:1309.4092 [hep-ph]} \BibitemShut
  {NoStop}%
\bibitem [{\citenamefont {Chambers}\ \emph {et~al.}(2014)\citenamefont
  {Chambers} \emph {et~al.}}]{Chambers:2014qaa}%
  \BibitemOpen
  \bibfield  {author} {\bibinfo {author} {\bibfnamefont {A.J.}\ \bibnamefont
  {Chambers}} \emph {et~al.} (\bibinfo {collaboration} {CSSM, QCDSF/UKQCD}),\
  }\bibfield  {title} {\enquote {\bibinfo {title} {{Feynman-Hellmann approach
  to the spin structure of hadrons}},}\ }\href {\doibase
  10.1103/PhysRevD.90.014510} {\bibfield  {journal} {\bibinfo  {journal}
  {Phys.Rev.}\ }\textbf {\bibinfo {volume} {D90}},\ \bibinfo {pages} {014510}
  (\bibinfo {year} {2014})},\ \Eprint {http://arxiv.org/abs/1405.3019}
  {arXiv:1405.3019 [hep-lat]} \BibitemShut {NoStop}%
\bibitem [{\citenamefont {Chambers}\ \emph {et~al.}(2015)\citenamefont
  {Chambers} \emph {et~al.}}]{Chambers:2015bka}%
  \BibitemOpen
  \bibfield  {author} {\bibinfo {author} {\bibfnamefont {A.~J.}\ \bibnamefont
  {Chambers}} \emph {et~al.},\ }\bibfield  {title} {\enquote {\bibinfo {title}
  {{Disconnected contributions to the spin of the nucleon}},}\ }\href {\doibase
  10.1103/PhysRevD.92.114517} {\bibfield  {journal} {\bibinfo  {journal} {Phys.
  Rev.}\ }\textbf {\bibinfo {volume} {D92}},\ \bibinfo {pages} {114517}
  (\bibinfo {year} {2015})},\ \Eprint {http://arxiv.org/abs/1508.06856}
  {arXiv:1508.06856 [hep-lat]} \BibitemShut {NoStop}%
\bibitem [{\citenamefont {Fucito}\ \emph {et~al.}(1982)\citenamefont {Fucito},
  \citenamefont {Parisi},\ and\ \citenamefont {Petrarca}}]{Fucito:1982ff}%
  \BibitemOpen
  \bibfield  {author} {\bibinfo {author} {\bibfnamefont {F.}~\bibnamefont
  {Fucito}}, \bibinfo {author} {\bibfnamefont {G.}~\bibnamefont {Parisi}}, \
  and\ \bibinfo {author} {\bibfnamefont {S.}~\bibnamefont {Petrarca}},\
  }\bibfield  {title} {\enquote {\bibinfo {title} {{FIRST EVALUATION OF G(A) /
  G(V) IN LATTICE QCD IN THE QUENCHED APPROXIMATION}},}\ }\href {\doibase
  10.1016/0370-2693(82)90816-4} {\bibfield  {journal} {\bibinfo  {journal}
  {Phys. Lett.}\ }\textbf {\bibinfo {volume} {B115}},\ \bibinfo {pages}
  {148--150} (\bibinfo {year} {1982})}\BibitemShut {NoStop}%
\bibitem [{\citenamefont {Martinelli}\ \emph {et~al.}(1982)\citenamefont
  {Martinelli}, \citenamefont {Parisi}, \citenamefont {Petronzio},\ and\
  \citenamefont {Rapuano}}]{Martinelli:1982cb}%
  \BibitemOpen
  \bibfield  {author} {\bibinfo {author} {\bibfnamefont {G.}~\bibnamefont
  {Martinelli}}, \bibinfo {author} {\bibfnamefont {G.}~\bibnamefont {Parisi}},
  \bibinfo {author} {\bibfnamefont {R.}~\bibnamefont {Petronzio}}, \ and\
  \bibinfo {author} {\bibfnamefont {F.}~\bibnamefont {Rapuano}},\ }\bibfield
  {title} {\enquote {\bibinfo {title} {{The Proton and Neutron Magnetic Moments
  in Lattice {QCD}}},}\ }\href {\doibase 10.1016/0370-2693(82)90162-9}
  {\bibfield  {journal} {\bibinfo  {journal} {Phys. Lett.}\ }\textbf {\bibinfo
  {volume} {B116}},\ \bibinfo {pages} {434--436} (\bibinfo {year}
  {1982})}\BibitemShut {NoStop}%
\bibitem [{\citenamefont {Bernard}\ \emph {et~al.}(1982)\citenamefont
  {Bernard}, \citenamefont {Draper}, \citenamefont {Olynyk},\ and\
  \citenamefont {Rushton}}]{Bernard:1982yu}%
  \BibitemOpen
  \bibfield  {author} {\bibinfo {author} {\bibfnamefont {Claude~W.}\
  \bibnamefont {Bernard}}, \bibinfo {author} {\bibfnamefont {Terrence}\
  \bibnamefont {Draper}}, \bibinfo {author} {\bibfnamefont {Kirk}\ \bibnamefont
  {Olynyk}}, \ and\ \bibinfo {author} {\bibfnamefont {Minick}\ \bibnamefont
  {Rushton}},\ }\bibfield  {title} {\enquote {\bibinfo {title} {{Lattice {QCD}
  Calculation of Some Baryon Magnetic Moments}},}\ }\href {\doibase
  10.1103/PhysRevLett.49.1076} {\bibfield  {journal} {\bibinfo  {journal}
  {Phys. Rev. Lett.}\ }\textbf {\bibinfo {volume} {49}},\ \bibinfo {pages}
  {1076} (\bibinfo {year} {1982})}\BibitemShut {NoStop}%
\bibitem [{\citenamefont {Detmold}\ \emph {et~al.}(2006)\citenamefont
  {Detmold}, \citenamefont {Tiburzi},\ and\ \citenamefont
  {Walker-Loud}}]{Detmold:2006vu}%
  \BibitemOpen
  \bibfield  {author} {\bibinfo {author} {\bibfnamefont {W.}~\bibnamefont
  {Detmold}}, \bibinfo {author} {\bibfnamefont {B.C.}\ \bibnamefont {Tiburzi}},
  \ and\ \bibinfo {author} {\bibfnamefont {Andre}\ \bibnamefont
  {Walker-Loud}},\ }\bibfield  {title} {\enquote {\bibinfo {title}
  {{Electromagnetic and spin polarisabilities in lattice QCD}},}\ }\href
  {\doibase 10.1103/PhysRevD.73.114505} {\bibfield  {journal} {\bibinfo
  {journal} {Phys.Rev.}\ }\textbf {\bibinfo {volume} {D73}},\ \bibinfo {pages}
  {114505} (\bibinfo {year} {2006})},\ \Eprint
  {http://arxiv.org/abs/hep-lat/0603026} {arXiv:hep-lat/0603026 [hep-lat]}
  \BibitemShut {NoStop}%
\bibitem [{\citenamefont {Engelhardt}(2007)}]{Engelhardt:2007ub}%
  \BibitemOpen
  \bibfield  {author} {\bibinfo {author} {\bibfnamefont {Michael}\ \bibnamefont
  {Engelhardt}} (\bibinfo {collaboration} {LHPC}),\ }\bibfield  {title}
  {\enquote {\bibinfo {title} {{Neutron electric polarizability from unquenched
  lattice QCD using the background field approach}},}\ }\href {\doibase
  10.1103/PhysRevD.76.114502} {\bibfield  {journal} {\bibinfo  {journal} {Phys.
  Rev.}\ }\textbf {\bibinfo {volume} {D76}},\ \bibinfo {pages} {114502}
  (\bibinfo {year} {2007})},\ \Eprint {http://arxiv.org/abs/0706.3919}
  {arXiv:0706.3919 [hep-lat]} \BibitemShut {NoStop}%
\bibitem [{\citenamefont {Detmold}\ \emph {et~al.}(2009)\citenamefont
  {Detmold}, \citenamefont {Tiburzi},\ and\ \citenamefont
  {Walker-Loud}}]{Detmold:2009dx}%
  \BibitemOpen
  \bibfield  {author} {\bibinfo {author} {\bibfnamefont {William}\ \bibnamefont
  {Detmold}}, \bibinfo {author} {\bibfnamefont {Brian~C.}\ \bibnamefont
  {Tiburzi}}, \ and\ \bibinfo {author} {\bibfnamefont {Andre}\ \bibnamefont
  {Walker-Loud}},\ }\bibfield  {title} {\enquote {\bibinfo {title} {{Extracting
  Electric Polarizabilities from Lattice QCD}},}\ }\href {\doibase
  10.1103/PhysRevD.79.094505} {\bibfield  {journal} {\bibinfo  {journal} {Phys.
  Rev.}\ }\textbf {\bibinfo {volume} {D79}},\ \bibinfo {pages} {094505}
  (\bibinfo {year} {2009})},\ \Eprint {http://arxiv.org/abs/0904.1586}
  {arXiv:0904.1586 [hep-lat]} \BibitemShut {NoStop}%
\bibitem [{\citenamefont {Detmold}\ \emph {et~al.}(2010)\citenamefont
  {Detmold}, \citenamefont {Tiburzi},\ and\ \citenamefont
  {Walker-Loud}}]{Detmold:2010ts}%
  \BibitemOpen
  \bibfield  {author} {\bibinfo {author} {\bibfnamefont {W.}~\bibnamefont
  {Detmold}}, \bibinfo {author} {\bibfnamefont {B.C.}\ \bibnamefont {Tiburzi}},
  \ and\ \bibinfo {author} {\bibfnamefont {A.}~\bibnamefont {Walker-Loud}},\
  }\bibfield  {title} {\enquote {\bibinfo {title} {{Extracting Nucleon Magnetic
  Moments and Electric Polarizabilities from Lattice QCD in Background Electric
  Fields}},}\ }\href {\doibase 10.1103/PhysRevD.81.054502} {\bibfield
  {journal} {\bibinfo  {journal} {Phys.Rev.}\ }\textbf {\bibinfo {volume}
  {D81}},\ \bibinfo {pages} {054502} (\bibinfo {year} {2010})},\ \Eprint
  {http://arxiv.org/abs/1001.1131} {arXiv:1001.1131 [hep-lat]} \BibitemShut
  {NoStop}%
\bibitem [{\citenamefont {Savage}\ \emph {et~al.}(2016)\citenamefont {Savage},
  \citenamefont {Shanahan}, \citenamefont {Tiburzi}, \citenamefont {Wagman},
  \citenamefont {Winter}, \citenamefont {Beane}, \citenamefont {Chang},
  \citenamefont {Davoudi}, \citenamefont {Detmold},\ and\ \citenamefont
  {Orginos}}]{Savage:2016kon}%
  \BibitemOpen
  \bibfield  {author} {\bibinfo {author} {\bibfnamefont {Martin~J.}\
  \bibnamefont {Savage}}, \bibinfo {author} {\bibfnamefont {Phiala~E.}\
  \bibnamefont {Shanahan}}, \bibinfo {author} {\bibfnamefont {Brian~C.}\
  \bibnamefont {Tiburzi}}, \bibinfo {author} {\bibfnamefont {Michael~L.}\
  \bibnamefont {Wagman}}, \bibinfo {author} {\bibfnamefont {Frank}\
  \bibnamefont {Winter}}, \bibinfo {author} {\bibfnamefont {Silas~R.}\
  \bibnamefont {Beane}}, \bibinfo {author} {\bibfnamefont {Emmanuel}\
  \bibnamefont {Chang}}, \bibinfo {author} {\bibfnamefont {Zohreh}\
  \bibnamefont {Davoudi}}, \bibinfo {author} {\bibfnamefont {William}\
  \bibnamefont {Detmold}}, \ and\ \bibinfo {author} {\bibfnamefont {Kostas}\
  \bibnamefont {Orginos}},\ }\bibfield  {title} {\enquote {\bibinfo {title}
  {{Proton-proton fusion and tritium $\beta$-decay from lattice quantum
  chromodynamics}},}\ }\href@noop {} {\  (\bibinfo {year} {2016})},\ \Eprint
  {http://arxiv.org/abs/1610.04545} {arXiv:1610.04545 [hep-lat]} \BibitemShut
  {NoStop}%
\bibitem [{\citenamefont {Basak}\ \emph
  {et~al.}(2005{\natexlab{a}})\citenamefont {Basak}, \citenamefont {Edwards},
  \citenamefont {Fleming}, \citenamefont {Heller}, \citenamefont {Morningstar},
  \citenamefont {Richards}, \citenamefont {Sato},\ and\ \citenamefont
  {Wallace}}]{Basak:2005aq}%
  \BibitemOpen
  \bibfield  {author} {\bibinfo {author} {\bibfnamefont {S.}~\bibnamefont
  {Basak}}, \bibinfo {author} {\bibfnamefont {R.~G.}\ \bibnamefont {Edwards}},
  \bibinfo {author} {\bibfnamefont {G.~T.}\ \bibnamefont {Fleming}}, \bibinfo
  {author} {\bibfnamefont {U.~M.}\ \bibnamefont {Heller}}, \bibinfo {author}
  {\bibfnamefont {C.}~\bibnamefont {Morningstar}}, \bibinfo {author}
  {\bibfnamefont {D.}~\bibnamefont {Richards}}, \bibinfo {author}
  {\bibfnamefont {I.}~\bibnamefont {Sato}}, \ and\ \bibinfo {author}
  {\bibfnamefont {S.}~\bibnamefont {Wallace}},\ }\bibfield  {title} {\enquote
  {\bibinfo {title} {{Group-theoretical construction of extended baryon
  operators in lattice QCD}},}\ }\href {\doibase 10.1103/PhysRevD.72.094506}
  {\bibfield  {journal} {\bibinfo  {journal} {Phys. Rev.}\ }\textbf {\bibinfo
  {volume} {D72}},\ \bibinfo {pages} {094506} (\bibinfo {year}
  {2005}{\natexlab{a}})},\ \Eprint {http://arxiv.org/abs/hep-lat/0506029}
  {arXiv:hep-lat/0506029 [hep-lat]} \BibitemShut {NoStop}%
\bibitem [{\citenamefont {Basak}\ \emph
  {et~al.}(2005{\natexlab{b}})\citenamefont {Basak}, \citenamefont {Edwards},
  \citenamefont {Fleming}, \citenamefont {Heller}, \citenamefont {Morningstar},
  \citenamefont {Richards}, \citenamefont {Sato},\ and\ \citenamefont
  {Wallace}}]{Basak:2005ir}%
  \BibitemOpen
  \bibfield  {author} {\bibinfo {author} {\bibfnamefont {Subhasish}\
  \bibnamefont {Basak}}, \bibinfo {author} {\bibfnamefont {Robert}\
  \bibnamefont {Edwards}}, \bibinfo {author} {\bibfnamefont {George~T.}\
  \bibnamefont {Fleming}}, \bibinfo {author} {\bibfnamefont {Urs~M.}\
  \bibnamefont {Heller}}, \bibinfo {author} {\bibfnamefont {Colin}\
  \bibnamefont {Morningstar}}, \bibinfo {author} {\bibfnamefont {David}\
  \bibnamefont {Richards}}, \bibinfo {author} {\bibfnamefont {Ikuro}\
  \bibnamefont {Sato}}, \ and\ \bibinfo {author} {\bibfnamefont {Stephen~J.}\
  \bibnamefont {Wallace}} (\bibinfo {collaboration} {Lattice Hadron Physics
  (LHPC)}),\ }\bibfield  {title} {\enquote {\bibinfo {title} {{Clebsch-Gordan
  construction of lattice interpolating fields for excited baryons}},}\ }\href
  {\doibase 10.1103/PhysRevD.72.074501} {\bibfield  {journal} {\bibinfo
  {journal} {Phys. Rev.}\ }\textbf {\bibinfo {volume} {D72}},\ \bibinfo {pages}
  {074501} (\bibinfo {year} {2005}{\natexlab{b}})},\ \Eprint
  {http://arxiv.org/abs/hep-lat/0508018} {arXiv:hep-lat/0508018 [hep-lat]}
  \BibitemShut {NoStop}%
\bibitem [{\citenamefont {Maiani}\ \emph {et~al.}(1987)\citenamefont {Maiani},
  \citenamefont {Martinelli}, \citenamefont {Paciello},\ and\ \citenamefont
  {Taglienti}}]{Maiani:1987by}%
  \BibitemOpen
  \bibfield  {author} {\bibinfo {author} {\bibfnamefont {L.}~\bibnamefont
  {Maiani}}, \bibinfo {author} {\bibfnamefont {G.}~\bibnamefont {Martinelli}},
  \bibinfo {author} {\bibfnamefont {M.~L.}\ \bibnamefont {Paciello}}, \ and\
  \bibinfo {author} {\bibfnamefont {B.}~\bibnamefont {Taglienti}},\ }\bibfield
  {title} {\enquote {\bibinfo {title} {{Scalar Densities and Baryon Mass
  Differences in Lattice {QCD} With Wilson Fermions}},}\ }\href {\doibase
  10.1016/0550-3213(87)90078-2} {\bibfield  {journal} {\bibinfo  {journal}
  {Nucl. Phys.}\ }\textbf {\bibinfo {volume} {B293}},\ \bibinfo {pages} {420}
  (\bibinfo {year} {1987})}\BibitemShut {NoStop}%
\bibitem [{\citenamefont {Bulava}\ \emph {et~al.}(2012)\citenamefont {Bulava},
  \citenamefont {Donnellan},\ and\ \citenamefont {Sommer}}]{Bulava:2011yz}%
  \BibitemOpen
  \bibfield  {author} {\bibinfo {author} {\bibfnamefont {John}\ \bibnamefont
  {Bulava}}, \bibinfo {author} {\bibfnamefont {Michael}\ \bibnamefont
  {Donnellan}}, \ and\ \bibinfo {author} {\bibfnamefont {Rainer}\ \bibnamefont
  {Sommer}},\ }\bibfield  {title} {\enquote {\bibinfo {title} {{On the
  computation of hadron-to-hadron transition matrix elements in lattice
  QCD}},}\ }\href {\doibase 10.1007/JHEP01(2012)140} {\bibfield  {journal}
  {\bibinfo  {journal} {JHEP}\ }\textbf {\bibinfo {volume} {01}},\ \bibinfo
  {pages} {140} (\bibinfo {year} {2012})},\ \Eprint
  {http://arxiv.org/abs/1108.3774} {arXiv:1108.3774 [hep-lat]} \BibitemShut
  {NoStop}%
\bibitem [{\citenamefont {Bernardoni}\ \emph {et~al.}(2015)\citenamefont
  {Bernardoni}, \citenamefont {Bulava}, \citenamefont {Donnellan},\ and\
  \citenamefont {Sommer}}]{Bernardoni:2014kla}%
  \BibitemOpen
  \bibfield  {author} {\bibinfo {author} {\bibfnamefont {Fabio}\ \bibnamefont
  {Bernardoni}}, \bibinfo {author} {\bibfnamefont {John}\ \bibnamefont
  {Bulava}}, \bibinfo {author} {\bibfnamefont {Michael}\ \bibnamefont
  {Donnellan}}, \ and\ \bibinfo {author} {\bibfnamefont {Rainer}\ \bibnamefont
  {Sommer}} (\bibinfo {collaboration} {ALPHA}),\ }\bibfield  {title} {\enquote
  {\bibinfo {title} {{Precision lattice QCD computation of the $B^*B\pi$
  coupling}},}\ }\href {\doibase 10.1016/j.physletb.2014.11.051} {\bibfield
  {journal} {\bibinfo  {journal} {Phys. Lett.}\ }\textbf {\bibinfo {volume}
  {B740}},\ \bibinfo {pages} {278--284} (\bibinfo {year} {2015})},\ \Eprint
  {http://arxiv.org/abs/1404.6951} {arXiv:1404.6951 [hep-lat]} \BibitemShut
  {NoStop}%
\bibitem [{\citenamefont {de~Divitiis}\ \emph {et~al.}(2012)\citenamefont
  {de~Divitiis}, \citenamefont {Petronzio},\ and\ \citenamefont
  {Tantalo}}]{deDivitiis:2012vs}%
  \BibitemOpen
  \bibfield  {author} {\bibinfo {author} {\bibfnamefont {G.~M.}\ \bibnamefont
  {de~Divitiis}}, \bibinfo {author} {\bibfnamefont {R.}~\bibnamefont
  {Petronzio}}, \ and\ \bibinfo {author} {\bibfnamefont {N.}~\bibnamefont
  {Tantalo}},\ }\bibfield  {title} {\enquote {\bibinfo {title} {{On the
  extraction of zero momentum form factors on the lattice}},}\ }\href {\doibase
  10.1016/j.physletb.2012.10.035} {\bibfield  {journal} {\bibinfo  {journal}
  {Phys. Lett.}\ }\textbf {\bibinfo {volume} {B718}},\ \bibinfo {pages}
  {589--596} (\bibinfo {year} {2012})},\ \Eprint
  {http://arxiv.org/abs/1208.5914} {arXiv:1208.5914 [hep-lat]} \BibitemShut
  {NoStop}%
\bibitem [{\citenamefont {Alexandrou}\ \emph {et~al.}(2014)\citenamefont
  {Alexandrou}, \citenamefont {Dinter}, \citenamefont {Drach}, \citenamefont
  {Jansen}, \citenamefont {Hadjiyiannakou},\ and\ \citenamefont
  {Renner}}]{Alexandrou:2013xon}%
  \BibitemOpen
  \bibfield  {author} {\bibinfo {author} {\bibfnamefont {Constantia}\
  \bibnamefont {Alexandrou}}, \bibinfo {author} {\bibfnamefont {Simon}\
  \bibnamefont {Dinter}}, \bibinfo {author} {\bibfnamefont {Vincent}\
  \bibnamefont {Drach}}, \bibinfo {author} {\bibfnamefont {Karl}\ \bibnamefont
  {Jansen}}, \bibinfo {author} {\bibfnamefont {Kyriakos}\ \bibnamefont
  {Hadjiyiannakou}}, \ and\ \bibinfo {author} {\bibfnamefont {Dru~B.}\
  \bibnamefont {Renner}} (\bibinfo {collaboration} {ETM}),\ }\bibfield  {title}
  {\enquote {\bibinfo {title} {{A Stochastic Method for Computing Hadronic
  Matrix Elements}},}\ }\href {\doibase 10.1140/epjc/s10052-013-2692-3}
  {\bibfield  {journal} {\bibinfo  {journal} {Eur. Phys. J.}\ }\textbf
  {\bibinfo {volume} {C74}},\ \bibinfo {pages} {2692} (\bibinfo {year}
  {2014})},\ \Eprint {http://arxiv.org/abs/1302.2608} {arXiv:1302.2608
  [hep-lat]} \BibitemShut {NoStop}%
\bibitem [{\citenamefont {Follana}\ \emph {et~al.}(2007)\citenamefont
  {Follana}, \citenamefont {Mason}, \citenamefont {Davies}, \citenamefont
  {Hornbostel}, \citenamefont {Lepage}, \citenamefont {Shigemitsu},
  \citenamefont {Trottier},\ and\ \citenamefont {Wong}}]{Follana:2006rc}%
  \BibitemOpen
  \bibfield  {author} {\bibinfo {author} {\bibfnamefont {E.}~\bibnamefont
  {Follana}}, \bibinfo {author} {\bibfnamefont {Q.}~\bibnamefont {Mason}},
  \bibinfo {author} {\bibfnamefont {C.}~\bibnamefont {Davies}}, \bibinfo
  {author} {\bibfnamefont {K.}~\bibnamefont {Hornbostel}}, \bibinfo {author}
  {\bibfnamefont {G.~P.}\ \bibnamefont {Lepage}}, \bibinfo {author}
  {\bibfnamefont {J.}~\bibnamefont {Shigemitsu}}, \bibinfo {author}
  {\bibfnamefont {H.}~\bibnamefont {Trottier}}, \ and\ \bibinfo {author}
  {\bibfnamefont {K.}~\bibnamefont {Wong}} (\bibinfo {collaboration} {HPQCD,
  UKQCD}),\ }\bibfield  {title} {\enquote {\bibinfo {title} {{Highly improved
  staggered quarks on the lattice, with applications to charm physics}},}\
  }\href {\doibase 10.1103/PhysRevD.75.054502} {\bibfield  {journal} {\bibinfo
  {journal} {Phys. Rev.}\ }\textbf {\bibinfo {volume} {D75}},\ \bibinfo {pages}
  {054502} (\bibinfo {year} {2007})},\ \Eprint
  {http://arxiv.org/abs/hep-lat/0610092} {arXiv:hep-lat/0610092 [hep-lat]}
  \BibitemShut {NoStop}%
\bibitem [{\citenamefont {Bazavov}\ \emph {et~al.}(2013)\citenamefont {Bazavov}
  \emph {et~al.}}]{Bazavov:2012xda}%
  \BibitemOpen
  \bibfield  {author} {\bibinfo {author} {\bibfnamefont {A.}~\bibnamefont
  {Bazavov}} \emph {et~al.} (\bibinfo {collaboration} {MILC}),\ }\bibfield
  {title} {\enquote {\bibinfo {title} {{Lattice QCD ensembles with four flavors
  of highly improved staggered quarks}},}\ }\href {\doibase
  10.1103/PhysRevD.87.054505} {\bibfield  {journal} {\bibinfo  {journal} {Phys.
  Rev.}\ }\textbf {\bibinfo {volume} {D87}},\ \bibinfo {pages} {054505}
  (\bibinfo {year} {2013})},\ \Eprint {http://arxiv.org/abs/1212.4768}
  {arXiv:1212.4768 [hep-lat]} \BibitemShut {NoStop}%
\bibitem [{\citenamefont {Brower}\ \emph {et~al.}(2012)\citenamefont {Brower},
  \citenamefont {Neff},\ and\ \citenamefont {Orginos}}]{Brower:2012vk}%
  \BibitemOpen
  \bibfield  {author} {\bibinfo {author} {\bibfnamefont {Richard~C.}\
  \bibnamefont {Brower}}, \bibinfo {author} {\bibfnamefont {Harmut}\
  \bibnamefont {Neff}}, \ and\ \bibinfo {author} {\bibfnamefont {Kostas}\
  \bibnamefont {Orginos}},\ }\bibfield  {title} {\enquote {\bibinfo {title}
  {{The M\'obius Domain Wall Fermion Algorithm}},}\ }\href@noop {} {\
  (\bibinfo {year} {2012})},\ \Eprint {http://arxiv.org/abs/1206.5214}
  {arXiv:1206.5214 [hep-lat]} \BibitemShut {NoStop}%
\bibitem [{\citenamefont {Clark}\ \emph {et~al.}(2010)\citenamefont {Clark},
  \citenamefont {Babich}, \citenamefont {Barros}, \citenamefont {Brower},\ and\
  \citenamefont {Rebbi}}]{Clark:2009wm}%
  \BibitemOpen
  \bibfield  {author} {\bibinfo {author} {\bibfnamefont {M.A.}\ \bibnamefont
  {Clark}}, \bibinfo {author} {\bibfnamefont {R.}~\bibnamefont {Babich}},
  \bibinfo {author} {\bibfnamefont {K.}~\bibnamefont {Barros}}, \bibinfo
  {author} {\bibfnamefont {R.C.}\ \bibnamefont {Brower}}, \ and\ \bibinfo
  {author} {\bibfnamefont {C.}~\bibnamefont {Rebbi}},\ }\bibfield  {title}
  {\enquote {\bibinfo {title} {{Solving Lattice QCD systems of equations using
  mixed precision solvers on GPUs}},}\ }\href {\doibase
  10.1016/j.cpc.2010.05.002} {\bibfield  {journal} {\bibinfo  {journal}
  {Comput.Phys.Commun.}\ }\textbf {\bibinfo {volume} {181}},\ \bibinfo {pages}
  {1517--1528} (\bibinfo {year} {2010})},\ \Eprint
  {http://arxiv.org/abs/0911.3191} {arXiv:0911.3191 [hep-lat]} \BibitemShut
  {NoStop}%
\bibitem [{\citenamefont {Babich}\ \emph {et~al.}(2011)\citenamefont {Babich},
  \citenamefont {Clark}, \citenamefont {Joo}, \citenamefont {Shi},
  \citenamefont {Brower} \emph {et~al.}}]{Babich:2011np}%
  \BibitemOpen
  \bibfield  {author} {\bibinfo {author} {\bibfnamefont {R.}~\bibnamefont
  {Babich}}, \bibinfo {author} {\bibfnamefont {M.A.}\ \bibnamefont {Clark}},
  \bibinfo {author} {\bibfnamefont {B.}~\bibnamefont {Joo}}, \bibinfo {author}
  {\bibfnamefont {G.}~\bibnamefont {Shi}}, \bibinfo {author} {\bibfnamefont
  {R.C.}\ \bibnamefont {Brower}},  \emph {et~al.},\ }\bibfield  {title}
  {\enquote {\bibinfo {title} {{Scaling Lattice QCD beyond 100 GPUs}},}\
  }\href@noop {} {\  (\bibinfo {year} {2011})},\ \Eprint
  {http://arxiv.org/abs/1109.2935} {arXiv:1109.2935 [hep-lat]} \BibitemShut
  {NoStop}%
\bibitem [{\citenamefont {Berkowitz}\ \emph
  {et~al.}(2017{\natexlab{a}})\citenamefont {Berkowitz}, \citenamefont
  {Bouchard}, \citenamefont {Chang}, \citenamefont {Clark}, \citenamefont
  {Joo}, \citenamefont {Kurth}, \citenamefont {Monahan}, \citenamefont
  {Nicholson}, \citenamefont {Orginos}, \citenamefont {Rinaldi}, \citenamefont
  {Vranas},\ and\ \citenamefont {Walker-Loud}}]{Berkowitz:2017opd}%
  \BibitemOpen
  \bibfield  {author} {\bibinfo {author} {\bibfnamefont {E.}~\bibnamefont
  {Berkowitz}}, \bibinfo {author} {\bibfnamefont {C.}~\bibnamefont {Bouchard}},
  \bibinfo {author} {\bibfnamefont {C.C}\ \bibnamefont {Chang}}, \bibinfo
  {author} {\bibfnamefont {M.A.}\ \bibnamefont {Clark}}, \bibinfo {author}
  {\bibfnamefont {B}~\bibnamefont {Joo}}, \bibinfo {author} {\bibfnamefont
  {T.}~\bibnamefont {Kurth}}, \bibinfo {author} {\bibfnamefont
  {C.}~\bibnamefont {Monahan}}, \bibinfo {author} {\bibfnamefont
  {A.}~\bibnamefont {Nicholson}}, \bibinfo {author} {\bibfnamefont
  {K.}~\bibnamefont {Orginos}}, \bibinfo {author} {\bibfnamefont
  {E.}~\bibnamefont {Rinaldi}}, \bibinfo {author} {\bibfnamefont
  {P.}~\bibnamefont {Vranas}}, \ and\ \bibinfo {author} {\bibfnamefont
  {A.}~\bibnamefont {Walker-Loud}},\ }\bibfield  {title} {\enquote {\bibinfo
  {title} {{M\"obius Domain-Wall fermions on gradient-flowed dynamical HISQ
  ensembles}},}\ }\href@noop {} {\  (\bibinfo {year} {2017}{\natexlab{a}})},\
  \Eprint {http://arxiv.org/abs/1701.07559} {arXiv:1701.07559 [hep-lat]}
  \BibitemShut {NoStop}%
\bibitem [{\citenamefont {Nicholson}\ \emph {et~al.}(2016)\citenamefont
  {Nicholson}, \citenamefont {Berkowitz}, \citenamefont {Chang}, \citenamefont
  {Clark}, \citenamefont {Joo}, \citenamefont {Kurth}, \citenamefont {Rinaldi},
  \citenamefont {Tiburzi}, \citenamefont {Vranas},\ and\ \citenamefont
  {Walker-Loud}}]{Nicholson:2016byl}%
  \BibitemOpen
  \bibfield  {author} {\bibinfo {author} {\bibfnamefont {Amy}\ \bibnamefont
  {Nicholson}}, \bibinfo {author} {\bibfnamefont {Evan}\ \bibnamefont
  {Berkowitz}}, \bibinfo {author} {\bibfnamefont {Chia~Cheng}\ \bibnamefont
  {Chang}}, \bibinfo {author} {\bibfnamefont {M.~A.}\ \bibnamefont {Clark}},
  \bibinfo {author} {\bibfnamefont {Balint}\ \bibnamefont {Joo}}, \bibinfo
  {author} {\bibfnamefont {Thorsten}\ \bibnamefont {Kurth}}, \bibinfo {author}
  {\bibfnamefont {Enrico}\ \bibnamefont {Rinaldi}}, \bibinfo {author}
  {\bibfnamefont {Brian}\ \bibnamefont {Tiburzi}}, \bibinfo {author}
  {\bibfnamefont {Pavlos}\ \bibnamefont {Vranas}}, \ and\ \bibinfo {author}
  {\bibfnamefont {Andre}\ \bibnamefont {Walker-Loud}},\ }\bibfield  {title}
  {\enquote {\bibinfo {title} {{Neutrinoless double beta decay from lattice
  QCD}},}\ }in\ \href
  {http://inspirehep.net/record/1481814/files/arXiv:1608.04793.pdf} {\emph
  {\bibinfo {booktitle} {{Proceedings, 34th International Symposium on Lattice
  Field Theory (Lattice 2016): Southampton, UK, July 24-30, 2016}}}}\ (\bibinfo
  {year} {2016})\ \Eprint {http://arxiv.org/abs/1608.04793} {arXiv:1608.04793
  [hep-lat]} \BibitemShut {NoStop}%
\bibitem [{\citenamefont {Lepage}\ \emph {et~al.}(2002)\citenamefont {Lepage},
  \citenamefont {Clark}, \citenamefont {Davies}, \citenamefont {Hornbostel},
  \citenamefont {Mackenzie}, \citenamefont {Morningstar},\ and\ \citenamefont
  {Trottier}}]{Lepage:2001ym}%
  \BibitemOpen
  \bibfield  {author} {\bibinfo {author} {\bibfnamefont {G.~P.}\ \bibnamefont
  {Lepage}}, \bibinfo {author} {\bibfnamefont {B.}~\bibnamefont {Clark}},
  \bibinfo {author} {\bibfnamefont {C.~T.~H.}\ \bibnamefont {Davies}}, \bibinfo
  {author} {\bibfnamefont {K.}~\bibnamefont {Hornbostel}}, \bibinfo {author}
  {\bibfnamefont {P.~B.}\ \bibnamefont {Mackenzie}}, \bibinfo {author}
  {\bibfnamefont {C.}~\bibnamefont {Morningstar}}, \ and\ \bibinfo {author}
  {\bibfnamefont {H.}~\bibnamefont {Trottier}},\ }\bibfield  {title} {\enquote
  {\bibinfo {title} {{Constrained curve fitting}},}\ }\bibfield  {booktitle}
  {\emph {\bibinfo {booktitle} {{Contents of lattice 2001 proceedings}}},\
  }\href {\doibase 10.1016/S0920-5632(01)01638-3} {\bibfield  {journal}
  {\bibinfo  {journal} {Nucl. Phys. Proc. Suppl.}\ }\textbf {\bibinfo {volume}
  {106}},\ \bibinfo {pages} {12--20} (\bibinfo {year} {2002})},\ \bibinfo
  {note} {[,12(2001)]},\ \Eprint {http://arxiv.org/abs/hep-lat/0110175}
  {arXiv:hep-lat/0110175 [hep-lat]} \BibitemShut {NoStop}%
\bibitem [{\citenamefont {Chen}\ \emph {et~al.}(2004)\citenamefont {Chen},
  \citenamefont {Dong}, \citenamefont {Draper}, \citenamefont {Horvath},
  \citenamefont {Liu}, \citenamefont {Mathur}, \citenamefont {Tamhankar},
  \citenamefont {Srinivasan}, \citenamefont {Lee},\ and\ \citenamefont
  {Zhang}}]{Chen:2004gp}%
  \BibitemOpen
  \bibfield  {author} {\bibinfo {author} {\bibfnamefont {Ying}\ \bibnamefont
  {Chen}}, \bibinfo {author} {\bibfnamefont {Shao-Jing}\ \bibnamefont {Dong}},
  \bibinfo {author} {\bibfnamefont {Terrence}\ \bibnamefont {Draper}}, \bibinfo
  {author} {\bibfnamefont {Ivan}\ \bibnamefont {Horvath}}, \bibinfo {author}
  {\bibfnamefont {Keh-Fei}\ \bibnamefont {Liu}}, \bibinfo {author}
  {\bibfnamefont {Nilmani}\ \bibnamefont {Mathur}}, \bibinfo {author}
  {\bibfnamefont {Sonali}\ \bibnamefont {Tamhankar}}, \bibinfo {author}
  {\bibfnamefont {Cidambi}\ \bibnamefont {Srinivasan}}, \bibinfo {author}
  {\bibfnamefont {Frank~X.}\ \bibnamefont {Lee}}, \ and\ \bibinfo {author}
  {\bibfnamefont {Jian-bo}\ \bibnamefont {Zhang}},\ }\bibfield  {title}
  {\enquote {\bibinfo {title} {{The Sequential empirical bayes method: An
  Adaptive constrained-curve fitting algorithm for lattice QCD}},}\ }\href@noop
  {} {\  (\bibinfo {year} {2004})},\ \Eprint
  {http://arxiv.org/abs/hep-lat/0405001} {arXiv:hep-lat/0405001 [hep-lat]}
  \BibitemShut {NoStop}%
\bibitem [{\citenamefont {Yoon}\ \emph {et~al.}(2017)\citenamefont {Yoon} \emph
  {et~al.}}]{Yoon:2016jzj}%
  \BibitemOpen
  \bibfield  {author} {\bibinfo {author} {\bibfnamefont {Boram}\ \bibnamefont
  {Yoon}} \emph {et~al.},\ }\bibfield  {title} {\enquote {\bibinfo {title}
  {{Isovector charges of the nucleon from 2+1-flavor QCD with clover
  fermions}},}\ }\href {\doibase 10.1103/PhysRevD.95.074508} {\bibfield
  {journal} {\bibinfo  {journal} {Phys. Rev.}\ }\textbf {\bibinfo {volume}
  {D95}},\ \bibinfo {pages} {074508} (\bibinfo {year} {2017})},\ \Eprint
  {http://arxiv.org/abs/1611.07452} {arXiv:1611.07452 [hep-lat]} \BibitemShut
  {NoStop}%
\bibitem [{\citenamefont {Bazavov}\ \emph {et~al.}(2016)\citenamefont {Bazavov}
  \emph {et~al.}}]{Bazavov:2016nty}%
  \BibitemOpen
  \bibfield  {author} {\bibinfo {author} {\bibfnamefont {A.}~\bibnamefont
  {Bazavov}} \emph {et~al.} (\bibinfo {collaboration} {Fermilab Lattice,
  MILC}),\ }\bibfield  {title} {\enquote {\bibinfo {title} {{$B^0_{(s)}$-mixing
  matrix elements from lattice QCD for the Standard Model and beyond}},}\
  }\href {\doibase 10.1103/PhysRevD.93.113016} {\bibfield  {journal} {\bibinfo
  {journal} {Phys. Rev.}\ }\textbf {\bibinfo {volume} {D93}},\ \bibinfo {pages}
  {113016} (\bibinfo {year} {2016})},\ \Eprint
  {http://arxiv.org/abs/1602.03560} {arXiv:1602.03560 [hep-lat]} \BibitemShut
  {NoStop}%
\bibitem [{\citenamefont {Lepage}(2016)}]{lsqfit}%
  \BibitemOpen
  \bibfield  {author} {\bibinfo {author} {\bibfnamefont {G.~Peter}\
  \bibnamefont {Lepage}},\ }\href {\doibase 10.5281/zenodo.60221} {\enquote
  {\bibinfo {title} {{\tt lsqfit} v8.1},}\ } (\bibinfo {year}
  {2016})\BibitemShut {NoStop}%
\bibitem [{\citenamefont {Bhattacharya}\ \emph {et~al.}(2016)\citenamefont
  {Bhattacharya}, \citenamefont {Cirigliano}, \citenamefont {Cohen},
  \citenamefont {Gupta}, \citenamefont {Lin},\ and\ \citenamefont
  {Yoon}}]{Bhattacharya:2016zcn}%
  \BibitemOpen
  \bibfield  {author} {\bibinfo {author} {\bibfnamefont {Tanmoy}\ \bibnamefont
  {Bhattacharya}}, \bibinfo {author} {\bibfnamefont {Vincenzo}\ \bibnamefont
  {Cirigliano}}, \bibinfo {author} {\bibfnamefont {Saul}\ \bibnamefont
  {Cohen}}, \bibinfo {author} {\bibfnamefont {Rajan}\ \bibnamefont {Gupta}},
  \bibinfo {author} {\bibfnamefont {Huey-Wen}\ \bibnamefont {Lin}}, \ and\
  \bibinfo {author} {\bibfnamefont {Boram}\ \bibnamefont {Yoon}},\ }\bibfield
  {title} {\enquote {\bibinfo {title} {{Axial, Scalar and Tensor Charges of the
  Nucleon from 2+1+1-flavor Lattice QCD}},}\ }\href {\doibase
  10.1103/PhysRevD.94.054508} {\bibfield  {journal} {\bibinfo  {journal} {Phys.
  Rev.}\ }\textbf {\bibinfo {volume} {D94}},\ \bibinfo {pages} {054508}
  (\bibinfo {year} {2016})},\ \Eprint {http://arxiv.org/abs/1606.07049}
  {arXiv:1606.07049 [hep-lat]} \BibitemShut {NoStop}%
\bibitem [{\citenamefont {Berkowitz}\ \emph
  {et~al.}(2017{\natexlab{b}})\citenamefont {Berkowitz} \emph
  {et~al.}}]{Berkowitz:2017gql}%
  \BibitemOpen
  \bibfield  {author} {\bibinfo {author} {\bibfnamefont {Evan}\ \bibnamefont
  {Berkowitz}} \emph {et~al.},\ }\bibfield  {title} {\enquote {\bibinfo {title}
  {{An accurate calculation of the nucleon axial charge with lattice QCD}},}\
  }\href@noop {} {\  (\bibinfo {year} {2017}{\natexlab{b}})},\ \Eprint
  {http://arxiv.org/abs/1704.01114} {arXiv:1704.01114 [hep-lat]} \BibitemShut
  {NoStop}%
\bibitem [{\citenamefont {Carlson}\ \emph {et~al.}(2017)\citenamefont
  {Carlson}, \citenamefont {Gorchtein},\ and\ \citenamefont
  {Vanderhaeghen}}]{Carlson:2016cii}%
  \BibitemOpen
  \bibfield  {author} {\bibinfo {author} {\bibfnamefont {Carl~E.}\ \bibnamefont
  {Carlson}}, \bibinfo {author} {\bibfnamefont {Mikhail}\ \bibnamefont
  {Gorchtein}}, \ and\ \bibinfo {author} {\bibfnamefont {Marc}\ \bibnamefont
  {Vanderhaeghen}},\ }\bibfield  {title} {\enquote {\bibinfo {title}
  {{Two-photon exchange correction to $2S-2P$ splitting in muonic $^3$He
  ions}},}\ }\href {\doibase 10.1103/PhysRevA.95.012506} {\bibfield  {journal}
  {\bibinfo  {journal} {Phys. Rev.}\ }\textbf {\bibinfo {volume} {A95}},\
  \bibinfo {pages} {012506} (\bibinfo {year} {2017})},\ \Eprint
  {http://arxiv.org/abs/1611.06192} {arXiv:1611.06192 [nucl-th]} \BibitemShut
  {NoStop}%
\bibitem [{\citenamefont {Rislow}\ and\ \citenamefont
  {Carlson}(2013)}]{Rislow:2013vta}%
  \BibitemOpen
  \bibfield  {author} {\bibinfo {author} {\bibfnamefont {Benjamin~C.}\
  \bibnamefont {Rislow}}\ and\ \bibinfo {author} {\bibfnamefont {Carl~E.}\
  \bibnamefont {Carlson}},\ }\bibfield  {title} {\enquote {\bibinfo {title}
  {{Modification of electromagnetic structure functions for the $\gamma Z$-box
  diagram}},}\ }\href {\doibase 10.1103/PhysRevD.88.013018} {\bibfield
  {journal} {\bibinfo  {journal} {Phys. Rev.}\ }\textbf {\bibinfo {volume}
  {D88}},\ \bibinfo {pages} {013018} (\bibinfo {year} {2013})},\ \Eprint
  {http://arxiv.org/abs/1304.8113} {arXiv:1304.8113 [hep-ph]} \BibitemShut
  {NoStop}%
\bibitem [{\citenamefont {Androic}\ \emph {et~al.}(2013)\citenamefont
  {Androic}, \citenamefont {Armstrong}, \citenamefont {Asaturyan},
  \citenamefont {Averett}, \citenamefont {Balewski}, \citenamefont {Beaufait},
  \citenamefont {Beminiwattha}, \citenamefont {Benesch}, \citenamefont
  {Benmokhtar}, \citenamefont {Birchall}, \citenamefont {Carlini},
  \citenamefont {Cates}, \citenamefont {Cornejo}, \citenamefont {Covrig},
  \citenamefont {Dalton}, \citenamefont {Davis}, \citenamefont {Deconinck},
  \citenamefont {Diefenbach}, \citenamefont {Dowd}, \citenamefont {Dunne},
  \citenamefont {Dutta}, \citenamefont {Duvall}, \citenamefont {Elaasar},
  \citenamefont {Falk}, \citenamefont {Finn}, \citenamefont {Forest},
  \citenamefont {Gaskell}, \citenamefont {Gericke}, \citenamefont {Grames},
  \citenamefont {Gray}, \citenamefont {Grimm}, \citenamefont {Guo},
  \citenamefont {Hoskins}, \citenamefont {Johnston}, \citenamefont {Jones},
  \citenamefont {Jones}, \citenamefont {Jones}, \citenamefont
  {Kargiantoulakis}, \citenamefont {King}, \citenamefont {Korkmaz},
  \citenamefont {Kowalski}, \citenamefont {Leacock}, \citenamefont {Leckey},
  \citenamefont {Lee}, \citenamefont {Lee}, \citenamefont {Lee}, \citenamefont
  {MacEwan}, \citenamefont {Mack}, \citenamefont {Magee}, \citenamefont
  {Mahurin}, \citenamefont {Mammei}, \citenamefont {Martin}, \citenamefont
  {McHugh}, \citenamefont {Meekins}, \citenamefont {Mei}, \citenamefont
  {Michaels}, \citenamefont {Micherdzinska}, \citenamefont {Mkrtchyan},
  \citenamefont {Mkrtchyan}, \citenamefont {Morgan}, \citenamefont {Myers},
  \citenamefont {Narayan}, \citenamefont {Ndukum}, \citenamefont {Nelyubin},
  \citenamefont {Nuruzzaman}, \citenamefont {van Oers}, \citenamefont {Opper},
  \citenamefont {Page}, \citenamefont {Pan}, \citenamefont {Paschke},
  \citenamefont {Phillips}, \citenamefont {Pitt}, \citenamefont {Poelker},
  \citenamefont {Rajotte}, \citenamefont {Ramsay}, \citenamefont {Roche},
  \citenamefont {Sawatzky}, \citenamefont {Seva}, \citenamefont {Shabestari},
  \citenamefont {Silwal}, \citenamefont {Simicevic}, \citenamefont {Smith},
  \citenamefont {Solvignon}, \citenamefont {Spayde}, \citenamefont {Subedi},
  \citenamefont {Subedi}, \citenamefont {Suleiman}, \citenamefont {Tadevosyan},
  \citenamefont {Tobias}, \citenamefont {Tvaskis}, \citenamefont {Waidyawansa},
  \citenamefont {Wang}, \citenamefont {Wells}, \citenamefont {Wood},
  \citenamefont {Yang}, \citenamefont {Young},\ and\ \citenamefont
  {Zhamkochyan}}]{PhysRevLett.111.141803}%
  \BibitemOpen
  \bibfield  {author} {\bibinfo {author} {\bibfnamefont {D.}~\bibnamefont
  {Androic}}, \bibinfo {author} {\bibfnamefont {D.~S.}\ \bibnamefont
  {Armstrong}}, \bibinfo {author} {\bibfnamefont {A.}~\bibnamefont
  {Asaturyan}}, \bibinfo {author} {\bibfnamefont {T.}~\bibnamefont {Averett}},
  \bibinfo {author} {\bibfnamefont {J.}~\bibnamefont {Balewski}}, \bibinfo
  {author} {\bibfnamefont {J.}~\bibnamefont {Beaufait}}, \bibinfo {author}
  {\bibfnamefont {R.~S.}\ \bibnamefont {Beminiwattha}}, \bibinfo {author}
  {\bibfnamefont {J.}~\bibnamefont {Benesch}}, \bibinfo {author} {\bibfnamefont
  {F.}~\bibnamefont {Benmokhtar}}, \bibinfo {author} {\bibfnamefont
  {J.}~\bibnamefont {Birchall}}, \bibinfo {author} {\bibfnamefont {R.~D.}\
  \bibnamefont {Carlini}}, \bibinfo {author} {\bibfnamefont {G.~D.}\
  \bibnamefont {Cates}}, \bibinfo {author} {\bibfnamefont {J.~C.}\ \bibnamefont
  {Cornejo}}, \bibinfo {author} {\bibfnamefont {S.}~\bibnamefont {Covrig}},
  \bibinfo {author} {\bibfnamefont {M.~M.}\ \bibnamefont {Dalton}}, \bibinfo
  {author} {\bibfnamefont {C.~A.}\ \bibnamefont {Davis}}, \bibinfo {author}
  {\bibfnamefont {W.}~\bibnamefont {Deconinck}}, \bibinfo {author}
  {\bibfnamefont {J.}~\bibnamefont {Diefenbach}}, \bibinfo {author}
  {\bibfnamefont {J.~F.}\ \bibnamefont {Dowd}}, \bibinfo {author}
  {\bibfnamefont {J.~A.}\ \bibnamefont {Dunne}}, \bibinfo {author}
  {\bibfnamefont {D.}~\bibnamefont {Dutta}}, \bibinfo {author} {\bibfnamefont
  {W.~S.}\ \bibnamefont {Duvall}}, \bibinfo {author} {\bibfnamefont
  {M.}~\bibnamefont {Elaasar}}, \bibinfo {author} {\bibfnamefont {W.~R.}\
  \bibnamefont {Falk}}, \bibinfo {author} {\bibfnamefont {J.~M.}\ \bibnamefont
  {Finn}}, \bibinfo {author} {\bibfnamefont {T.}~\bibnamefont {Forest}},
  \bibinfo {author} {\bibfnamefont {D.}~\bibnamefont {Gaskell}}, \bibinfo
  {author} {\bibfnamefont {M.~T.~W.}\ \bibnamefont {Gericke}}, \bibinfo
  {author} {\bibfnamefont {J.}~\bibnamefont {Grames}}, \bibinfo {author}
  {\bibfnamefont {V.~M.}\ \bibnamefont {Gray}}, \bibinfo {author}
  {\bibfnamefont {K.}~\bibnamefont {Grimm}}, \bibinfo {author} {\bibfnamefont
  {F.}~\bibnamefont {Guo}}, \bibinfo {author} {\bibfnamefont {J.~R.}\
  \bibnamefont {Hoskins}}, \bibinfo {author} {\bibfnamefont {K.}~\bibnamefont
  {Johnston}}, \bibinfo {author} {\bibfnamefont {D.}~\bibnamefont {Jones}},
  \bibinfo {author} {\bibfnamefont {M.}~\bibnamefont {Jones}}, \bibinfo
  {author} {\bibfnamefont {R.}~\bibnamefont {Jones}}, \bibinfo {author}
  {\bibfnamefont {M.}~\bibnamefont {Kargiantoulakis}}, \bibinfo {author}
  {\bibfnamefont {P.~M.}\ \bibnamefont {King}}, \bibinfo {author}
  {\bibfnamefont {E.}~\bibnamefont {Korkmaz}}, \bibinfo {author} {\bibfnamefont
  {S.}~\bibnamefont {Kowalski}}, \bibinfo {author} {\bibfnamefont
  {J.}~\bibnamefont {Leacock}}, \bibinfo {author} {\bibfnamefont
  {J.}~\bibnamefont {Leckey}}, \bibinfo {author} {\bibfnamefont {A.~R.}\
  \bibnamefont {Lee}}, \bibinfo {author} {\bibfnamefont {J.~H.}\ \bibnamefont
  {Lee}}, \bibinfo {author} {\bibfnamefont {L.}~\bibnamefont {Lee}}, \bibinfo
  {author} {\bibfnamefont {S.}~\bibnamefont {MacEwan}}, \bibinfo {author}
  {\bibfnamefont {D.}~\bibnamefont {Mack}}, \bibinfo {author} {\bibfnamefont
  {J.~A.}\ \bibnamefont {Magee}}, \bibinfo {author} {\bibfnamefont
  {R.}~\bibnamefont {Mahurin}}, \bibinfo {author} {\bibfnamefont
  {J.}~\bibnamefont {Mammei}}, \bibinfo {author} {\bibfnamefont {J.~W.}\
  \bibnamefont {Martin}}, \bibinfo {author} {\bibfnamefont {M.~J.}\
  \bibnamefont {McHugh}}, \bibinfo {author} {\bibfnamefont {D.}~\bibnamefont
  {Meekins}}, \bibinfo {author} {\bibfnamefont {J.}~\bibnamefont {Mei}},
  \bibinfo {author} {\bibfnamefont {R.}~\bibnamefont {Michaels}}, \bibinfo
  {author} {\bibfnamefont {A.}~\bibnamefont {Micherdzinska}}, \bibinfo {author}
  {\bibfnamefont {A.}~\bibnamefont {Mkrtchyan}}, \bibinfo {author}
  {\bibfnamefont {H.}~\bibnamefont {Mkrtchyan}}, \bibinfo {author}
  {\bibfnamefont {N.}~\bibnamefont {Morgan}}, \bibinfo {author} {\bibfnamefont
  {K.~E.}\ \bibnamefont {Myers}}, \bibinfo {author} {\bibfnamefont
  {A.}~\bibnamefont {Narayan}}, \bibinfo {author} {\bibfnamefont {L.~Z.}\
  \bibnamefont {Ndukum}}, \bibinfo {author} {\bibfnamefont {V.}~\bibnamefont
  {Nelyubin}}, \bibinfo {author} {\bibnamefont {Nuruzzaman}}, \bibinfo {author}
  {\bibfnamefont {W.~T.~H.}\ \bibnamefont {van Oers}}, \bibinfo {author}
  {\bibfnamefont {A.~K.}\ \bibnamefont {Opper}}, \bibinfo {author}
  {\bibfnamefont {S.~A.}\ \bibnamefont {Page}}, \bibinfo {author}
  {\bibfnamefont {J.}~\bibnamefont {Pan}}, \bibinfo {author} {\bibfnamefont
  {K.~D.}\ \bibnamefont {Paschke}}, \bibinfo {author} {\bibfnamefont {S.~K.}\
  \bibnamefont {Phillips}}, \bibinfo {author} {\bibfnamefont {M.~L.}\
  \bibnamefont {Pitt}}, \bibinfo {author} {\bibfnamefont {M.}~\bibnamefont
  {Poelker}}, \bibinfo {author} {\bibfnamefont {J.~F.}\ \bibnamefont
  {Rajotte}}, \bibinfo {author} {\bibfnamefont {W.~D.}\ \bibnamefont {Ramsay}},
  \bibinfo {author} {\bibfnamefont {J.}~\bibnamefont {Roche}}, \bibinfo
  {author} {\bibfnamefont {B.}~\bibnamefont {Sawatzky}}, \bibinfo {author}
  {\bibfnamefont {T.}~\bibnamefont {Seva}}, \bibinfo {author} {\bibfnamefont
  {M.~H.}\ \bibnamefont {Shabestari}}, \bibinfo {author} {\bibfnamefont
  {R.}~\bibnamefont {Silwal}}, \bibinfo {author} {\bibfnamefont
  {N.}~\bibnamefont {Simicevic}}, \bibinfo {author} {\bibfnamefont {G.~R.}\
  \bibnamefont {Smith}}, \bibinfo {author} {\bibfnamefont {P.}~\bibnamefont
  {Solvignon}}, \bibinfo {author} {\bibfnamefont {D.~T.}\ \bibnamefont
  {Spayde}}, \bibinfo {author} {\bibfnamefont {A.}~\bibnamefont {Subedi}},
  \bibinfo {author} {\bibfnamefont {R.}~\bibnamefont {Subedi}}, \bibinfo
  {author} {\bibfnamefont {R.}~\bibnamefont {Suleiman}}, \bibinfo {author}
  {\bibfnamefont {V.}~\bibnamefont {Tadevosyan}}, \bibinfo {author}
  {\bibfnamefont {W.~A.}\ \bibnamefont {Tobias}}, \bibinfo {author}
  {\bibfnamefont {V.}~\bibnamefont {Tvaskis}}, \bibinfo {author} {\bibfnamefont
  {B.}~\bibnamefont {Waidyawansa}}, \bibinfo {author} {\bibfnamefont
  {P.}~\bibnamefont {Wang}}, \bibinfo {author} {\bibfnamefont {S.~P.}\
  \bibnamefont {Wells}}, \bibinfo {author} {\bibfnamefont {S.~A.}\ \bibnamefont
  {Wood}}, \bibinfo {author} {\bibfnamefont {S.}~\bibnamefont {Yang}}, \bibinfo
  {author} {\bibfnamefont {R.~D.}\ \bibnamefont {Young}}, \ and\ \bibinfo
  {author} {\bibfnamefont {S.}~\bibnamefont {Zhamkochyan}} (\bibinfo
  {collaboration} {$Q_{\it weak}$ Collaboration}),\ }\bibfield  {title}
  {\enquote {\bibinfo {title} {First determination of the weak charge of the
  proton},}\ }\href {\doibase 10.1103/PhysRevLett.111.141803} {\bibfield
  {journal} {\bibinfo  {journal} {Phys. Rev. Lett.}\ }\textbf {\bibinfo
  {volume} {111}},\ \bibinfo {pages} {141803} (\bibinfo {year}
  {2013})}\BibitemShut {NoStop}%
\bibitem [{\citenamefont {Edwards}\ and\ \citenamefont
  {Joo}(2005)}]{Edwards:2004sx}%
  \BibitemOpen
  \bibfield  {author} {\bibinfo {author} {\bibfnamefont {Robert~G.}\
  \bibnamefont {Edwards}}\ and\ \bibinfo {author} {\bibfnamefont {Balint}\
  \bibnamefont {Joo}} (\bibinfo {collaboration} {SciDAC Collaboration, LHPC
  Collaboration, UKQCD Collaboration}),\ }\bibfield  {title} {\enquote
  {\bibinfo {title} {{The Chroma software system for lattice QCD}},}\ }\href
  {\doibase 10.1016/j.nuclphysbps.2004.11.254} {\bibfield  {journal} {\bibinfo
  {journal} {Nucl.Phys.Proc.Suppl.}\ }\textbf {\bibinfo {volume} {140}},\
  \bibinfo {pages} {832} (\bibinfo {year} {2005})},\ \Eprint
  {http://arxiv.org/abs/hep-lat/0409003} {arXiv:hep-lat/0409003 [hep-lat]}
  \BibitemShut {NoStop}%
\bibitem [{\citenamefont {{The HDF Group}}(1997-NNNN)}]{hdf5}%
  \BibitemOpen
  \bibfield  {author} {\bibinfo {author} {\bibnamefont {{The HDF Group}}},\
  }\href@noop {} {\enquote {\bibinfo {title} {{Hierarchical Data Format,
  version 5}},}\ } (\bibinfo {year} {1997-NNNN}),\ \bibinfo {note}
  {http://www.hdfgroup.org/HDF5/}\BibitemShut {NoStop}%
\bibitem [{\citenamefont {Kurth}\ \emph {et~al.}(2015)\citenamefont {Kurth},
  \citenamefont {Pochinsky}, \citenamefont {Sarje}, \citenamefont {Syritsyn},\
  and\ \citenamefont {Walker-Loud}}]{Kurth:2015mqa}%
  \BibitemOpen
  \bibfield  {author} {\bibinfo {author} {\bibfnamefont {Thorsten}\
  \bibnamefont {Kurth}}, \bibinfo {author} {\bibfnamefont {Andrew}\
  \bibnamefont {Pochinsky}}, \bibinfo {author} {\bibfnamefont {Abhinav}\
  \bibnamefont {Sarje}}, \bibinfo {author} {\bibfnamefont {Sergey}\
  \bibnamefont {Syritsyn}}, \ and\ \bibinfo {author} {\bibfnamefont {Andre}\
  \bibnamefont {Walker-Loud}},\ }\bibfield  {title} {\enquote {\bibinfo {title}
  {{High-Performance I/O: HDF5 for Lattice QCD}},}\ }\bibfield  {booktitle}
  {\emph {\bibinfo {booktitle} {{Proceedings, 32nd International Symposium on
  Lattice Field Theory (Lattice 2014): Brookhaven, NY, USA, June 23-28,
  2014}}},\ }\href@noop {} {\bibfield  {journal} {\bibinfo  {journal} {PoS}\
  }\textbf {\bibinfo {volume} {LATTICE2014}},\ \bibinfo {pages} {045} (\bibinfo
  {year} {2015})},\ \Eprint {http://arxiv.org/abs/1501.06992} {arXiv:1501.06992
  [hep-lat]} \BibitemShut {NoStop}%
\end{thebibliography}%

\end{document}